\documentclass[
groupedaddress,
preprint,
showkeys,
showpacs,
nofootinbib,
 amsmath,amssymb,
 aps,
pra,
]{revtex4-1}
\usepackage{amsmath,amssymb,amsthm,amsfonts,latexsym,graphicx,subfigure,multirow}
\usepackage{mathrsfs}
\usepackage[dvips]{color}
\usepackage[all,import]{xy}
\usepackage{indentfirst}
\usepackage{fancyhdr}
\usepackage{enumerate}
\usepackage{accents}
\usepackage{stmaryrd}

\usepackage{makeidx}                 %
\usepackage[hyperindex=true,
           pdfstartview=FitH,
           bookmarksnumbered=true,
           bookmarksopen=true,
           colorlinks=false, %
           pdfborder=001,   %
           citecolor=blue,%
           linkcolor=blue
           ]{hyperref}
\makeindex

\makeatletter
\def\wbar{\accentset{{\cc@style\underline{\mskip8mu}}}}
\makeatother

\makeatletter
\def\wbar{\accentset{{\cc@style\underline{\mskip8mu}}}}

\makeatother

\renewcommand{\vec}[1]{\mbox{\boldmath \small $#1$}}

\def\mi{\mathtt{i}}

\def\sech{\mathrm{sech}} 
\def\me{\mathrm{e}} 

\def\dif{\mathrm{d}}

\newcommand{\diag}{\mathrm{diag}}

\theoremstyle{definition}

\newcommand{\ie}{\textit{i.e.}{~}}
\newcommand{\eg}{\textit{e.g.}{~}}

\newcommand{\vgamma}{\vec{\gamma}}

\newcommand{\vPsi}{\vec{\Psi}}

\newcommand{\bq}{\begin{equation}}
\newcommand{\ba}{\begin{eqnarray}}
\newcommand{\eq}{\end{equation}}
\newcommand{\ea}{\end{eqnarray}}
\newcommand {\bPsi}{{\bar \Psi}}
\newcommand {\bpsi}{{\bar \psi}}
\usepackage{graphicx}

\begin{document}
\title{Stability of solitary waves  in the nonlinear Dirac equation with arbitrary nonlinearity}
\author{Sihong Shao}\email{sihong@math.pku.edu.cn}
\affiliation{LMAM and School of Mathematical Sciences, Peking University, Beijing 100871, China}
\author{Niurka R. Quintero}\email{niurka@us.es}
\affiliation{IMUS and Departamento de F{\'i}sica Aplicada I, E.S.P. Universidad de Sevilla, 41011 Sevilla, Spain}
\author{Franz G. Mertens}
\email{Franz.Mertens@uni-bayreuth.de}
\affiliation{Physikalisches Institut, Universit{\" a}t Bayreuth, D-95440 Bayreuth, Germany} 
\author{Fred Cooper} \email{cooper@santafe.edu}
\affiliation{Theoretical Division, Los Alamos National Laboratory,
Los Alamos, NM 87545, USA}
\affiliation{The Santa Fe Institute, 1399 Hyde Park Road, Santa Fe, NM 87501, USA}
\author{Avinash Khare} 
\email{khare@iiserpune.ac.in}
\affiliation{ Indian Institute of Science Education and Research, Pune 
411021, India}
\author{Avadh Saxena} \email{avadh@lanl.gov}
\affiliation{Theoretical Division and Center for Nonlinear Studies, Los Alamos National Laboratory,
Los Alamos, NM 87545, USA}

\date{\today}

\begin{abstract}
We consider the nonlinear Dirac equation in 1+1 dimension with scalar-scalar  self interaction $ \frac{ g^2}{ \kappa+1} ( {\bar \Psi} \Psi)^{ \kappa+1}$ and with mass $m$.    Using  the exact analytic form for  rest frame solitary waves of the form $\Psi(x,t) = \psi(x) e^{-i \omega t}$  for arbitrary $ \kappa$, we  
discuss the validity  of various approaches to understanding stability that were successful for the nonlinear Schr\"odinger equation. In particular we study the validity of a version of  Derrick's theorem, the criterion of Bogolubsky  as well as  the Vakhitov-Kolokolov criterion, and find that these criteria yield inconsistent results. 
Therefore, we study the stability by numerical  simulations using a recently developed 4th-order 
operator splitting integration method. For different ranges of $\kappa$ we map 
out the stability regimes in $\omega$. 
We find that all stable nonlinear Dirac solitary waves have a one-hump profile, but not all one-hump waves are stable, while all waves with two humps are unstable.
We also find that the time $t_c$, it takes for the instability to set in, is an exponentially increasing function of $\omega$ and  
$t_c$ decreases monotonically with increasing $\kappa$.   
  \end{abstract}
\pacs{PACS: 11.15.Kc, 03.70.+ k, 0570.Ln.,11.10.-s}
\maketitle

\section{Introduction} \label{sec1}

The nonlinear Dirac equation has been studied \cite{ref:Lee} \cite{ref:Nogami} in detail in the past  for the particular case that  the nonlinearity parameter $ \kappa=1$  (massive Gross Neveu \cite{ref:GN}  and massive Thirring models \cite{ref:TM}). In those studies it was found that these equations have solitary wave solutions.  These solutions are of the form  $\Psi(x,t)= e^{-i\omega t} \psi(x)$ in the rest frame, where $\psi(x)$ is a 2-component spinor.  In a recent paper \cite{NLDE}  we generalized these solutions to arbitrary nonlinearity $ \kappa$ and compared the exact solutions with  the non-relativistic reduction of these solutions.  At that time there were conflicting statements about the stability of these solutions as to whether Bogolubsky's approach \cite{ref:bogol} for determining stability was valid. He suggested 
two approaches, one a variation of Derrick's theorem \cite{ref:derrick} which looks at stability with respect to scale transformations and suggested that for $\kappa >1$ the solitary wave should be unstable.  This
approach seemed to violate the continuity argument that the nonlinear Dirac (NLD) equation becomes a modified 
nonlinear Schr\"odinger (NLS) equation when $\omega$ approaches the mass parameter $m$ of the Dirac equation.  This argument has been made more rigorous by Comech \cite{comech}.  Comech (private communiction)  has been able to prove that  for
$ \kappa <2$, the Vakhitov-Kolokolov  \cite{stab1} criterion guarantees linear stability in the non-relativistic regime of the NLD equation for solutions of the form (in the rest frame) 
$\Psi(x,t) = \psi(x) e^{-i \omega t}$ where $\omega$ is less than but approximately equal to  $m$. 
He was also able to show linear instability in the same non-relativistic regime for $\kappa > 2$.  This is the first rigorous result for the Dirac equation that applies  in the non-relativistic regime.  Below when we refer to NLS or NLD, it would be implicit that we refer to these equations with arbitrary nonlinearity ($\kappa$).  

Bogolubsky also proposed another test for determining stability based on varying the frequency $\omega$, while keeping the charge fixed. In his paper \cite{ref:bogol}, Bogolubsky only used this approach for $\kappa=1$, since he believed that only at  $\kappa = 1$  did the stability argument based on scale transformations not apply. That argument  (which we will discuss in Section \ref{sec4}),  predicts  that for $\kappa <1 $ the solitary waves were stable under scale transformations and for $\kappa > 1$ they should be unstable to scale transformations. 
 This approach  for studying stability based on varying the frequency when extended to all values of $\kappa 
 \leq 2$  predicts that 
when $\omega \lesssim 0.7$ that the solitary waves should be unstable to changes in $\omega$ for fixed charge.  We also  show that the $\omega$ variational approach of Bogolubsky  is equivalent to assuming that  instability will occur in variational trial functions which preserve charge as we change $\omega$.   Finally we will discuss the  Vakhitov-Kolokolov  \cite{stab1} criterion as applied to the nonlinear Dirac equation.  We will show that it predicts for all $\kappa < 2$ that the solitary waves are stable for all values of $\omega$ and that there is a regime in $\omega$ even for $\kappa >2$ where the solitary waves are predicted to be linearly stable. However, these predictions are \textit{not} confirmed 
by our simulations (Section \ref{sec5}) which means that the Vakhitov-Kolokolov 
criterion is not valid for the NLD case.  Before applying these methods to the NLD equation, we show that these three variational approaches to stability all give the same result when applied to the NLS equation, namely for all
values of $\omega$ when $\kappa <2$ the solutions are stable, and for $\kappa > 2$ they are unstable. 

Previous studies of instability have been confined to the case $\kappa=1$.
Bogolubsky \cite{ref:bogol} studied this problem numerically after suggesting that solitary waves  of the nonlinear Dirac equation should be unstable 
if $\omega< \omega_{B} \approx 1/\sqrt{2}$ for $g=1$ and $m=1$. 
He presented in his paper results for 
$\omega = 0.5$  (unstable) and $\omega =0.8$ (stable)  but the integration times were not given.
In contrast to this, Alvarez and Soler \cite{alvarez} 
claimed based on their simulations that the solitary wave solutions for $\kappa=1$  were stable for 
all $\omega$ values. 
In our simulations, shown in the subsequent tables and figures we find that for $\omega < \omega_c $ the solitary waves are metastable with a lifetime $t_c$  growing exponentially below the $\omega_c$.  

The integration times in \cite{alvarez} are much too 
small to observe the instabilities we have found for $\omega < \omega_c$ . This also holds for the scattering 
experiments of  \cite{ref:numerical} which studied the collision of  two solitary waves with 
$\omega = 0.6$ and $0.8$ at $\kappa=1$. Here the former solitary wave looks stable, 
but the integration time is only about 100.  The simulations we have performed here have confirmed Bogolubsky's intuition that there is a critical value of $\omega$ below which the solitary waves are unstable, but they do not agree with his determination of the critical value.  Our 
simulations are in agreement with Comech's proof \cite{comech} that in the non-relativistic regime solitary waves should be stable for $\kappa < 2$,
and unstable for $\kappa > 2$.  

Our paper is organized as follows: in Section \ref{sec2} we review the exact solution for arbitrary $\kappa$. In Section \ref{sec3} we consider the non-relativistic limit which is the nonlinear  Schr{\" o}dinger equation with a linear mass term. We discuss all three variational methods as applied to the NLS equation, namely Derrick's Theorem, stability with respect to changes in $\omega$ for fixed charge, and the  Vakhitov-Kolokolov criterion.

 In Section \ref{sec4} we discuss how these three approaches when applied naively lead to different conclusions for the NLD equation. A version of Derrick's theorem predicts that all solitary waves with $\kappa >1 $ are unstable, which disagrees with Comech's results 
 \cite{comech} in the non-relativistic limit.
 Bogolubsky's criterion predicts that for $\omega$ less than a critical value, and $\kappa < 2$, the solutions should be unstable, but
 in the non-relativistic regime predicts stability.  Vakhitov-Kolokolov instead predicts all solutions should be stable for $\kappa < 2$ and there is a domain of stability for $\omega $ smaller than a critical value where again the solution should be stable for $\kappa > 2$.    In Section \ref{sec5} we present the results of detailed simulations of the nonlinear Dirac equation for $\kappa=1$, 
 $0<\kappa<1$,  $1<\kappa<2$, and $\kappa \geq 2$ and map out the stability regimes in $\omega$. For $0<\kappa \leq 1$ there is a stability regime for 
 $\omega_c \leq \omega < 1$, where the critical value $\omega_c$ increases monotonically with $\kappa$. For $1<\kappa<2$ there are two types of stability regions. For $\kappa \geq 2$ small stable regions exist, but only for $\kappa=2$ and values 
 slightly larger than $2$. 
 
 We also find for $\omega < \omega_c$ that the time $t_c$ it takes for the instability to set in is an exponentially increasing function of the frequency 
 $\omega$ and $t_c$ as a function of $\kappa$ decreases monotonically with 
 increasing $\kappa$. Moreover, we find that below $\kappa=2$ there is a non-relativistic regime of $\omega$ close to $m$ where the solitary waves are always stable. Finally, we remark that all stable NLD solitary waves have a one-hump profile, but not all one-hump waves are stable. All waves with two humps are unstable. Our conclusions are presented in Section 
 \ref{sec6}. 
 
\section{review of exact solutions} \label{sec2}

The NLD equations that we are interested in are given by
\bq
(i \gamma^{\mu} \partial_{\mu} - m) \Psi +g^2 (\bPsi  \Psi)^{ \kappa} \Psi = 0 , 
\eq
which can be derived in a standard fashion from the Lagrangian density

\bq
\mathcal{L} =  \left(\frac{i}{2}\right) [\bPsi \gamma^{\mu} \partial_{\mu} \Psi 
-\partial_{\mu} \bPsi \gamma^{\mu} \Psi] - m \bPsi \Psi +  {\cal L}_I ; ~~{\mathcal L}_I  =   \frac{g^2}{ \kappa+1} (\bPsi \Psi)^{ \kappa+1} \ 
\label{eq:t112} . 
\eq
For solitary wave solutions,  the  field $\Psi$ goes to zero at $x \to \pm \infty$.    It is sufficient to go into 
the rest frame to discuss the solutions, since the theory is Lorentz invariant and the moving solution can be obtained by a Lorentz boost.
In the rest frame we assume the wave function is of the form
\bq
\Psi(x,t) = e^{-i\omega t} \psi(x).  \label{restframe}
\eq
We are interested in bound state solutions that correspond to positive energy $\omega \geq 0$ and which have energies in the rest frame less than the mass parameter $m$, i.e. $\omega < m$. 
In our previous paper \cite{NLDE}, we chose the representation  $\gamma_0 = \sigma_3$ and $i \gamma_1= \sigma_1$. Here instead, to make contact with the numerical simulation paper of Alvarez and Carreras \cite{ref:numerical} we instead choose the representation
$ \gamma^0 = \sigma_3$ and $\gamma^1= i \sigma_2$. 

Defining the functions  $u(x)$, $v(x)$, $R(x)$, $\theta(x)$ via:
\ba
\psi(x) &&  =  \left(  \begin{array} {cc}
      u(x) \\
      i ~v(x) \\ 
   \end{array} \right) =
R(x) \left(\begin{array}{c}\cos \theta \\ i \sin \theta \end{array}\right),
\ea
we obtain the following equations for $u$ and $v$:
\ba \label{dirac2}
&& \frac{du}{dx} + (m+\omega ) v - g^2(u^2-v^2)^{ \kappa} v=0, \nonumber \\
&&\frac{dv}{dx} + (m-\omega ) u - g^2(u^2-v^2)^{ \kappa} u=0. \label{dirac21}
\ea
From energy-momentum conservation
\bq 
\partial^\mu  T_{\mu \nu} =0; \qquad   
 T_{\mu \nu} = \frac{i}{2} \left[ \bPsi \gamma_\mu \partial_ \nu \Psi  -  \partial_\nu \bPsi  \gamma_\mu  \Psi \right]  - g_{\mu \nu} {\cal L},
 \eq
we obtain  in the rest frame for stationary solutions
\bq  \label{con}
T_{10} = constant; \qquad T_{11} = constant.
\eq
Using (\ref{restframe}) we obtain
\bq
T_{11} = \omega  \psi^\dag \psi - m  \bpsi \psi + {\cal L}_I. 
\eq
For solitary wave solutions vanishing at $x \to \pm \infty$  the constant in Eq. (\ref{con})  is zero and we obtain
\bq
T_{11}= \omega  \psi^\dag \psi - m  \bpsi \psi + {\cal L}_I = 0.  \label{eq:t11}
\eq
Multiplying the  equation of motion  on the left by $\bPsi$  we have that
\bq
( \kappa+1)  {\cal L}_I = - \omega  \psi^\dag \psi + m  \bpsi \psi  + \bpsi i \gamma_1 \partial_1 \psi.  \label{eq:motion}
\eq
Therefore we can rewrite $T_{11} = 0$ as 
\bq
\omega   \kappa  \psi^\dag \psi - m  \kappa    \bpsi \psi  + \bpsi i \gamma_1 \partial_1 \psi = 0.
\eq
For  the Hamiltonian density we have
\bq
 {\cal H} =T_{00} =   \bPsi i \gamma_1 \partial_1 \Psi+ m \bPsi \Psi - {\cal L}_I  \equiv  h_1+ h_2- h_3. \label{eq:hdensity}
\eq
Each of ${h_i}$ are positive definite.
From Eq. (\ref{eq:t11}) and (\ref{eq:motion}) one has the relationship:
\bq
  \kappa  {\cal L}_I =   \bpsi i \gamma_1 \partial_1 \psi.    \label{eq:rel}
\eq
From this we have
\bq
h_3 = \frac{1} { \kappa}  h_1,   \label{eq:relation}
\eq
and in particular for $\kappa=1$, ${\cal H} = m \bpsi \psi$.
In terms of $R$, $\theta$
one has 
\bq
 \bpsi i \gamma_1 \partial_1 \psi = \psi^\dag \psi \frac{d \theta} {dx}.
 \eq
 This leads to the simple differential equation for $\theta$  for solitary waves 
 \bq
 \frac{d \theta} {dx} = -  \omega _ \kappa+ m_ \kappa \cos 2 \theta  ; ~~  
 ~~~\omega _ \kappa  \equiv  \kappa~ \omega; ~~ m_ \kappa =  \kappa~m.
 \eq
The solution, choosing the origin of the solitary wave to be at $x=0$ (which we will do in what follows), is
\bq
\theta(x) =  \tan^{-1} ( \alpha \tanh \beta_{ \kappa} x),
\eq
where 
\bq
\alpha = \left( \frac{m_ \kappa - \omega _ \kappa}{m_ \kappa + \omega _ \kappa} \right) ^{1/2}
=  \left( \frac{m - \omega }{m + \omega } \right) ^{1/2}, ~~ \beta_{ \kappa} = (m_ \kappa^2 - \omega _ \kappa^2)^{1/2}.
\eq
Thus we have
\ba \label{ident2}
\tan \theta(x)  &&=  \alpha \tanh \beta_{ \kappa} x , \nonumber \\
\sin^2 \theta(x) && =\frac{\alpha^2 \tanh^2 \beta_{ \kappa} x}{1+ \alpha^2 \tanh^2 \beta_{ \kappa} x} = \frac{(m-\omega) \sinh^2 
\beta_\kappa x}{m \cosh 2 \beta_{\kappa} x + \omega} ,  \nonumber \\
\cos^2 \theta(x) &&= \frac{1}{1+ \alpha^2 \tanh^2 \beta_{ \kappa} x}=  \frac{(m+\omega) \cosh^2 \beta_{\kappa} x}{m \cosh 2 \beta_{\kappa} x + \omega}, 
\ea
where we have used the identities:
\ba 
\label{ident1}
1+ \alpha^2 \tanh^2 \beta_{k} x & = & \left(\frac {m \cosh 2 \beta_{k} x + \omega }{m+\omega } \right) \sech^2\beta_{k} x  \>,
\nonumber \\
1- \alpha^2 \tanh^2 \beta_{k} x &  = & \left(\frac {\omega \cosh 2 \beta_{k} x + m}{m+\omega } \right)  \sech^2\beta_{k} x \>.
\ea
From 
 (\ref{eq:t112})  and (\ref{eq:t11})  we find
 \bq
R^2 =\left[  \frac{( \kappa+1) (m \cos 2 \theta -\omega ) }{g^2 (\cos 2 \theta)^{ \kappa+1} }\right] ^{1/ \kappa}.
\eq
Now we have
\bq
\frac{d \theta}{dx} = \frac{\beta_{ \kappa}^2}{\omega _ \kappa+m_ \kappa \cosh 2\beta_{ \kappa} x} = -\omega _ \kappa + m_ \kappa \cos 2 \theta,  \label{eq:dthetadx}
\eq
so that 
\bq  \cos 2 \theta =  \frac{m_ \kappa+\omega _ \kappa \cosh 2 \beta_{ \kappa} x}{\omega _ \kappa+m_ \kappa \cosh 2 \beta_{ \kappa} x }=  \frac{m+\omega  \cosh 2 \beta_{ \kappa} x}{\omega +m \cosh 2 \beta_{ \kappa} x }.  
\label{eq:cos2theta}  
\eq
One important expression is 
\bq
m \cos 2 \theta - \omega  = \frac {\beta_{ \kappa}^2} { \kappa^2(\omega +m \cosh 2 \beta_{ \kappa} x)}.
\eq
Using this we get
\bq
R^2 = \left( \frac {\omega +m \cosh 2 \beta_{ \kappa} x}{ m+\omega  \cosh 2 \beta_{ \kappa} x} \right)    \left[ \frac {( \kappa+1) \beta_{ \kappa}^2} 
{g^2  \kappa^2 (m+\omega  \cosh 2 \beta_{ \kappa} x)} \right]^{1/ \kappa}.
\eq
Using the identities of Eq. (\ref{ident1})
we obtain the alternate expression
\bq
R^2 =\left( \frac  {1+\alpha^2 \tanh^2\beta_{ \kappa} x }  {1-\alpha^2 \tanh^2\beta_{ \kappa} x } \right) 
  \left[ \frac{{\rm sech}^2\beta_{ \kappa} x  ( \kappa+1) \beta_{ \kappa}^2}{g^2 (m+\omega) \kappa^2 ( 1-\alpha^2 \tanh^2\beta_{ \kappa} x )} \right ]^{1/ \kappa}.  \label{eq:Rsq}
\eq
In particular for $\kappa=1$
\ba
R^2&&= \frac{2 (m-\omega)}{g^2}  \frac{(1+\alpha^2 \tanh^2 \beta x)}{(1-\alpha^2 \tanh^2 \beta x)^2} \sech^2 \beta x \nonumber \\
&& = \frac{2 \beta^2}{g^2} 
\frac{ (\omega+ m \cosh 2 \beta x)}{(m+\omega \cosh 2 \beta x)^2}.
\ea
Using the second equation for $R^2$ and Eq. (\ref{ident2})
we obtain 
\ba
u^2&&= R^2 \cos^2 \theta =  \frac{2}{g^2} \frac{(m^2-\omega^2) (m+\omega) \cosh^2 \beta x}{(m+\omega \cosh 2 \beta x)^2}, \nonumber \\
v^2&&= R^2 \sin^2 \theta =  \frac{2} {g^2} \frac{(m^2-\omega^2) (m-\omega) \sinh^2 \beta x}{(m+\omega \cosh 2 \beta x)^2}, 
\ea
which agrees with the expression in Alvarez and Carreras \cite{ref:numerical} with a redefinition of the coupling to our convention. 
For arbitrary $\kappa$ we have
   \ba
u^2 && =  \frac{(m+\omega)  \cosh ^2(\kappa \beta x)}{m+\omega \cosh(2\kappa \beta x)} 
\bigg [\frac{(\kappa+1) \beta ^2}{g^2 (m+\omega  \cosh (2\kappa \beta x))} 
\bigg ]^{\frac{1}{\kappa}}, \nonumber   \\ 
v^2 && =   \frac{(m-\omega) \sinh^2(\kappa \beta x)}{m+\omega \cosh(2\kappa \beta x)}
\bigg [\frac{(\kappa +1) \beta ^2}{g^2 (m+\omega  \cosh (2\kappa \beta x))} 
\bigg ]^{\frac{1}{\kappa}}. 
   \ea
The equation for $\omega$ in terms of $g^2$ is determined from the fact that the single solitary wave has charge Q. 
We have
\bq
 Q =    \int _{-\infty}^{+\infty} \psi^\dag \psi dx= \int _{-\infty}^{+\infty} dx R^2(x)  =\frac{1}{ \beta_{ \kappa} }\left[\frac{( \kappa+1) \beta_{ \kappa}^2}{g^2  \kappa^2 (m+\omega )}\right]^{1/\kappa} I_ \kappa[\alpha^2],  
  \label{eq:Q}
\eq
where
\ba
I_ \kappa[\alpha^2] 
&&=\int^{+1}_{-1} dy \frac{1+\alpha^2 y^2}{(1-y^2)^{( \kappa-1)/ \kappa}
[1-\alpha^2 y^2]^{( \kappa+1)/ \kappa}}. \nonumber \\
&& =B(1/2,1/ \kappa) _2F_1(1+1/ \kappa,1/2,1/2+1/ \kappa;\alpha^2)+ \nonumber \\
&&+\alpha^2 B(3/2,1/ \kappa) _2F_1(1+1/ \kappa,3/2,3/2+1/ \kappa;\alpha^2) ,   
\ea 
and $ _2F_1$ is a hypergeometric function and $B(p,q)$ is the beta function, also called the Eulerian integral of the first kind. 

To find $\omega$ as a function of $g^2$ and $Q$ one solves the equation
\bq
I_ \kappa[\alpha^2] = Q \beta_{ \kappa} \left[\frac{g^2  \kappa^2 (m+\omega )}{( \kappa+1) \beta_{ \kappa}^2}\right]^{1/\kappa}.
\eq
In what follows we will scale all parameters in terms of $m$ (i.e $\omega \rightarrow \omega/m $, etc.).
For $\kappa=1$, $Q$ has a very simple form  
\bq
Q=\int_{-\infty}^{+\infty} dx R^2 = \frac{4\alpha}{(1-\alpha^2)g^2}= \frac {2 \beta}{g^2 \omega}=\frac{2 \sqrt{1-\omega^2}}{g^2 \omega}. \label{eq:Q1} 
\eq
   
Now for $H_1 = \int_{-\infty}^{+\infty} dx~ R^2 d \theta/dx$ we have
\bq
H_1 =\frac{1}{ \kappa}  \left[\frac{ \beta_ \kappa^2}{m+\omega}\right ] ^{1+1/ \kappa}  \left[\frac{ \kappa+1}{ \kappa^2g^2} \right]^{1/ \kappa}  \int_{-\infty}^\infty dx \left[\frac{\sech^2\beta_ \kappa x}{1- \alpha^2 \tanh^2 \beta_ \kappa x} \right]^{1+1/ \kappa}.
\eq

Again changing variables, letting $y = \tanh \beta_ \kappa x$, we obtain:
\ba
H_1&& = \frac{1}{ \kappa \beta_ \kappa}  \left[\frac{ \beta_ \kappa^2}{m+\omega}\right ] ^{1+1/ \kappa}  \left[\frac{ \kappa+1}{ \kappa^2g^2} \right]^{1/ \kappa} J_ \kappa[\alpha], \nonumber \\
J_ \kappa[\alpha] && =
 \int_{-1}^1 dy~ \frac{(1-y^2)^{1/ \kappa}}{(1-\alpha^2 y^2)^{1+1/ \kappa}} \nonumber \\
&& =
 B\left(\frac{1}{2},1+\frac{1}{\kappa }\right) \,
   _2F_1\left(\frac{1}{2},1+\frac{1}{\kappa };\frac{3}{2}+\frac{1}{\kappa
   };\frac{1-\omega }{\omega +1}\right).
\ea

For $ \kappa=1$, 
\bq
H_1[ \kappa=1] = -\frac{2 \left(\sqrt{1-\omega^2}-2 \tanh
   ^{-1}\left(\sqrt{\frac{1-\omega}{\omega+1}}\right)\right)}{g^2}.
\eq

   Now for $H_2 = m \int_{-\infty}^\infty dx~ R^2 cos 2 \theta $ we have  
\bq
H_2 =    \left[\frac{( \kappa+1)}{ \kappa^2g^2(1+\omega)} \right]^{1/ \kappa}  \int_{-\infty}^\infty dx \left[\frac{\sech^2\beta_ \kappa x}{1- \alpha^2 \tanh^2 \beta_ \kappa x} \right]^{1/ \kappa}.
\eq

Again changing variables, letting $y = \tanh \beta_ \kappa x$, we obtain:
\ba
H_2&& =  \left[\frac{( \kappa+1)}{ \kappa^2g^2(1+\omega)} \right]^{1/ \kappa}  \left[  \beta_ \kappa^2 \right ] ^{1/ \kappa-1/2}   K_ \kappa[\alpha] , \nonumber \\
K_ \kappa[\alpha] &&=
 \int_{-1}^1 dy~ \frac{(1-y^2)^{1/ \kappa-1}}{(1-\alpha^2 y^2)^{1/ \kappa}} \nonumber \\
 &&=B\left(\frac{1}{2},\frac{1}{\kappa }\right) \,
   _2F_1\left(\frac{1}{2},\frac{1}{\kappa };\frac{1}{2}+\frac{1}{\kappa
   };\frac{1-\omega }{\omega +1}\right).
   \ea
At 
 $ \kappa=1$, 
\bq
H_2 = \frac{4 \tanh ^{-1}\left(\sqrt{\frac{1-\omega }{\omega+1}}\right)}{g^2}.
\eq
\section{The non-relativistic limit--Nonlinear Schr{\"o}dinger equation } \label{sec3}
In a previous paper  \cite{NLDE}  we showed that if we write the rest frame solutions as in Eqs.   (\ref{dirac2})-(\ref{dirac21}) and take the non-relativistic  limit where   $(m-\omega)/(2m) \ll 1$, then $u(x)$ obeys the equation:
\bq 
\omega ~ u(x)= -\frac{1}{2m} \frac{\partial^2}{\partial x^2}~ u(x)   + m ~u(x) - g^2  (u)^{2\kappa+1} .
\eq
Defining $\psi(x,t) = u(x) e^{-i \omega t}$ we find that $\psi(x,t)$ obeys 
nonlinear  Schr{\"o}dinger equation with a linear term proportional to $m$:
\bq
 i \frac{\partial}{\partial t} \psi + \frac{1}{2m} \frac{\partial^2}{\partial x^2} \psi + {g^2 }(\psi^\star \psi)^{ \kappa} \psi- m \psi  = 0,   \label{psieq}
\eq
(here $\hbar = c=1$,  but we keep the explicit dependence of $m$ for clarity in this section).
This equation has solutions of the form:  $\psi(x,t) = e^{-i \omega  t} \psi_\omega(x)$  where: 
\bq
\psi_\omega (x)  = A ~ \sech ^{1/\kappa}  \left [ \beta_k x  \right],  \label{exact}
\eq
and 
\bq
A^{2 \kappa} 
= \frac{\beta_k^2 (\kappa+1)}{2 m g^2 \kappa^2},
\eq
and $\omega$ is given by
\bq
\omega = m - \frac{\beta_{k}^2} {2m \kappa^2}.
\eq
Thus
\bq 
\beta_k =  \kappa \sqrt{2 m}  \sqrt{m- \omega}.
\eq
Note that the expression for $A^2$ can be obtained from Eq. (\ref{eq:Rsq}) for $R^2$ by letting $\alpha^2 \rightarrow 0$ and $m+\omega \rightarrow 2m$
and again in the expression for
\bq
\beta_{dirac} = \kappa \sqrt{m-\omega}  \sqrt{m+\omega}
\eq 
by replacing $m+\omega \rightarrow 2m$. 

The analogue of the ``charge" (as well as the non-relativistic limit of $Q$  in the Dirac equation) is the ``Mass''  given by
\ba
M[\omega] &&= \int  dx \psi_\omega ^\star \psi_\omega  =  \frac{A^2}{\beta_{k}}   \frac{\sqrt{\pi } \Gamma \left(\frac{1}{\kappa }\right)}{\Gamma \left(\frac{1}{2}+\frac{1}{\kappa} \right)}  \nonumber \\
&& = \left(\frac{\beta_k^2 (\kappa+1)}{2 m g^2 \kappa^2}\right) ^{1/\kappa}  \frac{1}{\beta_k}  \frac{\sqrt{\pi } \Gamma \left(\frac{1}{\kappa }\right)}{\Gamma \left(\frac{1}{2}+\frac{1}{\kappa }\right)}  
=\frac{\sqrt{\frac{\pi }{2}} \Gamma \left(\frac{1}{\kappa}\right) \left(\frac{(\kappa+1) (m-\omega
   )}{g^2}\right)^{\frac{1}{\kappa}}}{\kappa \sqrt{m} \sqrt{m-\omega} \Gamma
   \left(\frac{1}{2}+\frac{1}{\kappa}\right)}.
\ea
\subsection{Derrick's Theorem}
For the  NLS equation we can use the scaling argument of 
Derrick \cite{ref:derrick}  to determine if the solutions are unstable to scale transformation.  The Hamiltonian  is given by 
\bq
H = \int dx  \left \{\frac{1}{2m} \psi^\star_x \psi_x +m   \psi^\star \psi
-\frac{g^2}{\kappa+1} (\psi^\star \psi)^{\kappa+1} \right\}.
\eq
From the equations of motion one can show that when we evaluate $H$ for solitary wave solutions then  $ H_3 = \frac{2}{\kappa} H_1$.

Thus the value of the energy of a solitary wave solution  is given by
\bq
H= m M[\psi_\omega]  + \frac{\kappa-2}{2} H_3[\psi_\omega].
\eq
Here 
\ba
H_3 &&= \frac{g^2}{\kappa+1} \frac{A^{2 \kappa+2}}{\beta_k} \frac{\sqrt{\pi } \Gamma \left(1+\frac{1}{\kappa}\right)}{\Gamma
   \left(\frac{3}{2}+\frac{1}{\kappa}\right)} \nonumber \\
  &&= \frac{
  \sqrt{\frac{\pi (m-\omega)}{2 m}}  
  \Gamma \left(1+\frac{1}{\kappa}\right)
   \left(\frac{(\kappa+1) (m-\omega)}{g^2}\right)^\frac{1}{\kappa}} 
   {\kappa \Gamma \left(\frac{3}{2}+\frac{1}{\kappa}\right)}. 
\ea   
It is well known that using stability with respect to scale transformation to 
understand domains of stability applies to this type of Hamiltonian. 
This  Hamiltonian can be written 
\bq 
H= H_1 +m  H_2- H_3.
\eq
where $H_i >0$ $(i=1,2,3)$.  
If we  make a scale transformation on the solution which preserves the 
mass $M = \int  \psi^\star \psi dx$,
\bq
\psi_\lambda \rightarrow \lambda^{1/2} \psi(\lambda x),
\eq
we obtain
\bq
H_\lambda =  \lambda^2  H_1+m  H_2-  \lambda^\kappa  H_3.
\eq
The first derivative is
\bq
\frac{\partial H}{\partial \lambda} = 2 \lambda H_1 - \kappa  \lambda^{ \kappa-1}  H_3.
\eq
Setting the derivative to zero at $\lambda =1 $ gives the equation consistent 
with the equations of motion:
\bq
\kappa H_3 = 2 H_1.  
\eq
The second derivative at $\lambda =1$ can now be written as
\bq
\frac{\partial^2 H}{ \partial \lambda^2} =   \kappa (2- \kappa) H_3[\psi_\omega].    \label{Derrick2}
\eq
The solution is therefore unstable to scale transformations when $\kappa >2$. 

\subsection{Linear Stability and the Vakhitov-Kolokokov criterion} \label{sec3.2}

In the case of the nonlinear  Schr{\" o}dinger equation, it is easy to perform a linear stability analysis for the exact solutions.
Namely one lets
\bq
\psi(x,t) = \left( \psi_\omega(x) + r(x,t) \right)e^{-i \omega t},
\eq
linearizes the equation for $r(x,t)$
\bq
\partial_t r(x,t)  = A_\omega r(x,t),  
\eq
and studies the eigenvalues of the differential operator $A_\omega$.  If the spectrum of $A_\omega$ is imaginary, then the solutions are spectrally stable.  Vakhitov and Kolokolov \cite{stab1}  showed that when the spectrum is purely imaginary,  $dM[\omega]/d \omega <0$. 
Also they showed that when $dM[\omega]/d \omega >0$, there is a real positive  eigenvalue so that there is a linear instability.
For the NLS equation we have that
\bq
M[\omega] = k  \beta_{k}^{(2- \kappa) /\kappa} = k  (m-\omega) ^{(2- \kappa) /(2 \kappa)}   , ~~ k >0, 
\eq
where $k$ is positive real. 
Thus 
\bq
\frac{dM}{d \omega}  =  k' (\kappa-2); ~~ k' >0. 
\eq
Thus for $\kappa > 2$ the solitary waves are unstable.  

\subsection{Stability to changes in the frequency at fixed charge} \label{sec3.3}
In this section we will study the suggestion of Bogolubsky that we can determine stability by looking at whether the energy of the solitary wave is increased or decreased as we vary the frequency $\omega$ for fixed values of the charge.  
 That is if we parametrize a rest frame solitary wave solution of the NLS equation,  which has a charge $M[\omega]$, given by
\bq
\psi_s(x,t) = \chi_s(x,\omega) e^{-i\omega t},
\eq
then we choose our slightly changed wave function to be
\ba
\tilde{\psi}[ x,t, \omega', \omega ] &&= \frac{\sqrt{M[\omega]}}{\sqrt{ M[\omega']}}  \chi_s(x,\omega') e^{-i\omega' t}  \nonumber \\
&& \equiv f(\omega',\omega)  \chi_s(x,\omega') e^{-i\omega' t}  .
\ea
Then the wave function $\tilde{\psi} [x,t,\omega',\omega] $ has the same charge  as $\psi[x,t, \omega]$.  Inserting this wave function into the Hamiltonian
we get a new Hamiltonian $H_p$  depending on both $\omega', \omega$.  As a function of $\omega'$ the probe Hamiltonian $H_p$  is stationary   at the 
value $\omega'=\omega$.    The probe Hamiltonian has the form
\bq
H_p[\omega', \omega] = H_3[\omega'] \left( \frac{ \kappa}{2}  f(\omega',\omega)^2 -  f(\omega',\omega)^{2( \kappa+1)} \right) + m M[\omega'] f(\omega',\omega) ^2.
\eq

For this probe, the first derivative is identically zero for the exact solution when $\omega'=\omega$.  The second derivative with respect to 
$\omega'$  evaluated at  $\omega'=\omega$ is exactly zero at $\kappa=2$, 
it is then positive for all $\omega$  for  $\kappa < 2$ and  strictly negative for all  $\omega$  for $\kappa > 2$.   Thus this test agrees with all the other variational methods in giving instability for all  $\omega$ when $\kappa > 2$.  It has nothing to say at the critical value $\kappa =2$.    

The second derivative evaluated at $\omega' = \omega$ is explicitly given by 
\bq
{H_p}_{\omega'\omega'}|_{\omega'=\omega} = \frac{\sqrt{\pi } (2-\kappa ) (\kappa +1)^{\frac{1}{\kappa }} (m-\omega
   )^{\frac{1}{\kappa }-1} \Gamma \left(1+\frac{1}{\kappa }\right)}{4 \sqrt{2-2 \omega
   } \Gamma \left(\frac{3}{2}+\frac{1}{\kappa }\right)}.
   \eq

\section{Variational approaches to the Stability of Exact Solutions of the 
nonlinear Dirac Equation} \label{sec4}
In this section we will investigate whether we can extend the variational methods that were successful in determining the domain of stability in the non-relativistic regime could be extended to the full relativistic regime ($\omega < m$) of the NLD equation. We will see that 
these three approaches suggest totally different answers as to the domain of stability as a function of $\omega$.  

\subsection{Stability to scale transformations at fixed charge}\label{sec4.1}
The first approach to stability, originally due to Derrick \cite{ref:derrick}  was to look at how the solitary wave responds to a scale transformation. 
The argument goes as follows \cite{ref:bogol}.  Consider the scale transformation $x \rightarrow \lambda x$.
We will assume that  an exact solution minimizes   $H_\lambda$ when $\lambda =1$  with the constraint that 
the charge is kept fixed.   One then assumes that if the second derivative is negative at $\lambda =1$ then the solutions are unstable to 
scale transformations and thus unstable.  For the NLS equation, we showed in \cite{NLDE} that this argument led to the {\it same} criterion as the linear stability result \cite{stab1} that for $\kappa > 2$ the solitary waves are unstable.  

Bogolubsky applied this argument to the Dirac equation and obtained a  result, which we will present, that suggests that for the NLD equation for
$\kappa> 1 $ the solitary waves are unstable.  This disagrees with our intuition, presented in \cite{NLDE}  that in the non-relativistic regime the NLD solitary waves should obey the same pattern of instability as the NLS equation.  This intuition has been given more credence in the recent
linear stability analysis of the NLD equation by Comech \cite{comech} which relies on studying the NLD equation in the non-relativistic regime. 
In that study, it was found that in the non-relativistic regime, the stability of the NLD  equation solitary waves should go over to the NLS 
equation result that for $\kappa < 2$ the solitary waves are stable.  Our numerical evidence supports this analysis.

The solution is of the form 
\bq
\psi(x) =  \left(\begin{array}{c}u \\v\end{array}\right)  = R(x) \left(\begin{array}{c}\cos \theta \\ i \sin \theta \end{array}\right) e^{-i\omega t} \>.
\eq
If we want to keep the charge fixed we consider the following stretched solution:
\bq
\psi_\lambda(x) =  \left(\begin{array}{c}u \\v\end{array}\right)  =  \lambda^{\frac{1}{2}}  R(\lambda x) \left(\begin{array}{c}\cos \theta(\lambda x) \\ i \sin \theta(\lambda x)  \end{array}\right) e^{-i\omega t} \>.
\eq
The value of the Hamiltonian
\ba
H &=& \int dx \  \Bigl [ \bpsi i \gamma^1 \partial_1 \psi + m \bpsi \psi  - \frac {g^2}{\kappa+1} (\bpsi \psi)^{\kappa+1}  \Bigr ]
\nonumber \\
&\equiv& H_1 +H_2 - H_3,
\ea
for the  stretched solution is 
\bq
H_\lambda = \lambda H_1 + H_2 - \lambda^\kappa H_3,
\eq
where again $H_i$ are all positive definite. 
The first derivative is 
\bq
\frac{\partial H_\lambda}{\partial \lambda} = H_1- \kappa \lambda^{\kappa-1} H_3.
\eq
At the minimum,  setting $\lambda =1$ we find in general 
\bq
H_3 = \frac{1}{\kappa} H_1 , 
\eq
which is consistent with the equation of motion result we obtained earlier, see Eq. (\ref{eq:relation}). We see that for $\kappa=1$ the energy is given by just  $H_2$. 
The second derivative yields:
\bq
\frac{\partial^2 H_\lambda}{\partial \lambda^2} = - \kappa(\kappa-1) \lambda^{\kappa-2} H_3.
\eq
From this we see that if $\kappa >1$, this analysis would suggest that  solitary waves are unstable to small  changes in the width.  For $\kappa< 1$ the solitary waves are stable to this type of perturbation. The case $\kappa=1$  would require a separate treatment since this analysis yields no information.
This argument does not depend on ${\cal L}_I$ as long as ${\cal L}_I$ is positive definite. The weakness in this argument is that one needs to prove 
that the stable solutions of the NLD equation are not merely stationary solutions of the variational principle but are actually minima of 
$H_\lambda$. The fact that this idea disagrees both with the continuity argument of Comech \cite{comech} and our simulations makes us seriously doubt this assumption.   We find that  even at $\kappa =2$ there is a range of $\omega$ near $m$ where the solitary waves are stable.

\subsection{Stability to changes in the frequency at fixed charge} \label{sec4.2}
Bogolubsky   \cite{ref:bogol} suggested that the stability could be ascertained by looking at variations of the wave function,  keeping the charge fixed and seeing if the solution was a minimum  or maximum   of the Hamiltonian as a function of the parameter $\omega$.  If the deformed solution decreases the energy, then he assumed that this is a sufficient condition for the solitary wave to be unstable. Bobolubsky applied this criterion  for the case $\kappa=1$ since he presumably thought that Derrick's theorem was applicable at all other values of $\kappa$.  As we showed previously, this criterion agrees with all the other variational methods when applied to the NLS equation, with the Mass taking the place of the Charge when we study the NLS equation.  
 Assuming we  know the wave function at  the value of $\omega$ corresponding to a fixed charge $Q$,   if we change the parametric dependence on $\omega$ this also changes the charge.
This can be corrected by assuming that the new wave function has a new normalization that corrects for this.
That is if we parametrize a rest frame solitary wave solution of the NLD equation,  which has a charge $Q[\omega] $ given by
\bq
\psi_s(x,t) = \chi_s(x,\omega) e^{-i\omega t},
\eq
then we choose our slightly changed wave function to be
\ba
\tilde{\psi}[ x,t, \omega', \omega ] &&= \frac{\sqrt{Q[\omega]}}{\sqrt{ Q[\omega']}}  \chi_s(x,\omega') e^{-i\omega' t}  \nonumber \\
&& \equiv f(\omega',\omega)  \chi_s(x,\omega') e^{-i\omega' t}  .
\ea
Then the wave function $\tilde{\psi} [x,t,\omega',\omega] $ has the same charge as $\psi[x,t, \omega]$.  Inserting this wave function into the Hamiltonian
we get a new Hamiltonian $H_p$  depending on both $\omega', \omega$.  As a function of $\omega'$ the probe Hamiltonian $H_p$  is stationary   at the 
value $\omega'=\omega$.  The criterion Bogolubsky proposes is that the solitary wave is  unstable to this type of perturbation if  the  probe Hamiltonian has a maximum at $\omega' = \omega$.   What we will find using this approach is that the second derivative of the probe Hamiltonian is negative below  a critical value of $\omega$,  where $\omega_B \approx 0.7$, suggesting an instability for all $\omega$ less than this value.  For 
$\kappa \leq 2$ using this criterion we find a regime near $\omega = m$ where  $\omega < m$ and the second derivative is positive, suggesting stability
in the nonrelatistic regime in agreement with Comech \cite{comech}.   We will use the notation $\omega_B$ for the critical value of $\omega$ below which the Bogolubsky criterion leads to instability.

\begin{figure}[b]
    \centering
    \includegraphics[width=0.8\columnwidth]{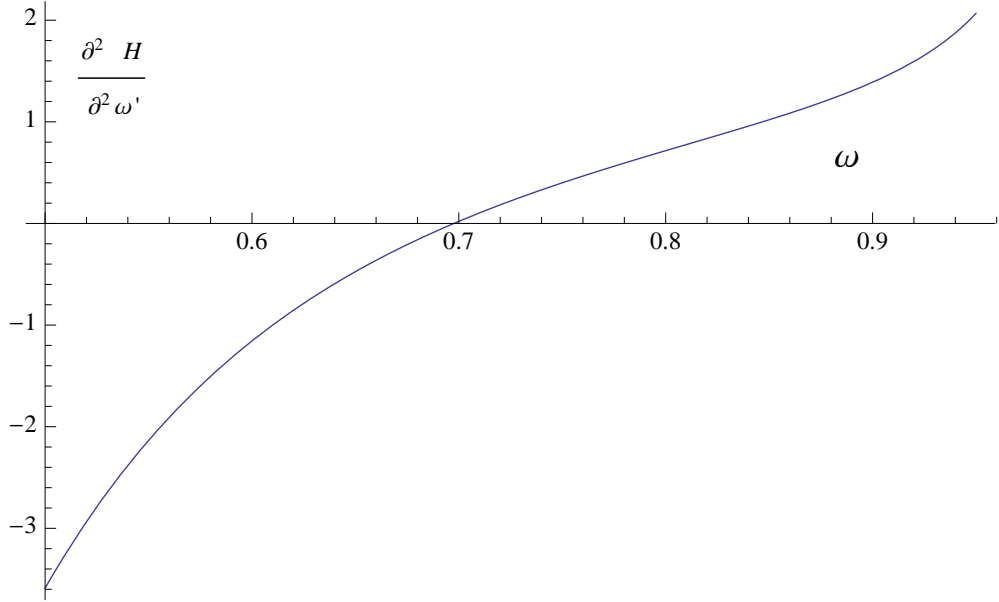}
    \caption{\label{one}(Color online)
    Second derivative of probe Hamiltonian at $\omega'=\omega$ as a function of $\omega$ for $\kappa = 1$.}
    \end{figure}

The probe Hamiltonian has the form:
\bq
H_p[\omega', \omega] = H_1[\omega'] \left( f(\omega',\omega)^2 - \frac{1}{ \kappa}  f(\omega',\omega)^{2( \kappa+1)} \right) + H_2[\omega'] f(\omega',\omega) ^2.
\eq
In what follows we will suppress the dependence of $H_p$ on $g$ since that dependence is multiplicative, namely $H_p \propto \frac{1}{g^{2 \kappa}} $. 
For all values of $\kappa$ we find that the first derivative of $H_p$ with respect to $\omega'$ evaluated at $\omega' = \omega$ is indeed zero.  The behavior of the second derivative evaluated at $\omega' = \omega$ as a function of  $\omega$, is different as we change $\kappa$.  For $\kappa <2$  the second derivative becomes negative for $\omega < \omega_B  \approx 0.7$ and then becomes positive 
above that value.  This is seen in Fig. \ref{one} for $\kappa=1$. 


\begin{figure}[b]  
    \centering
    \includegraphics[width=0.8\columnwidth]{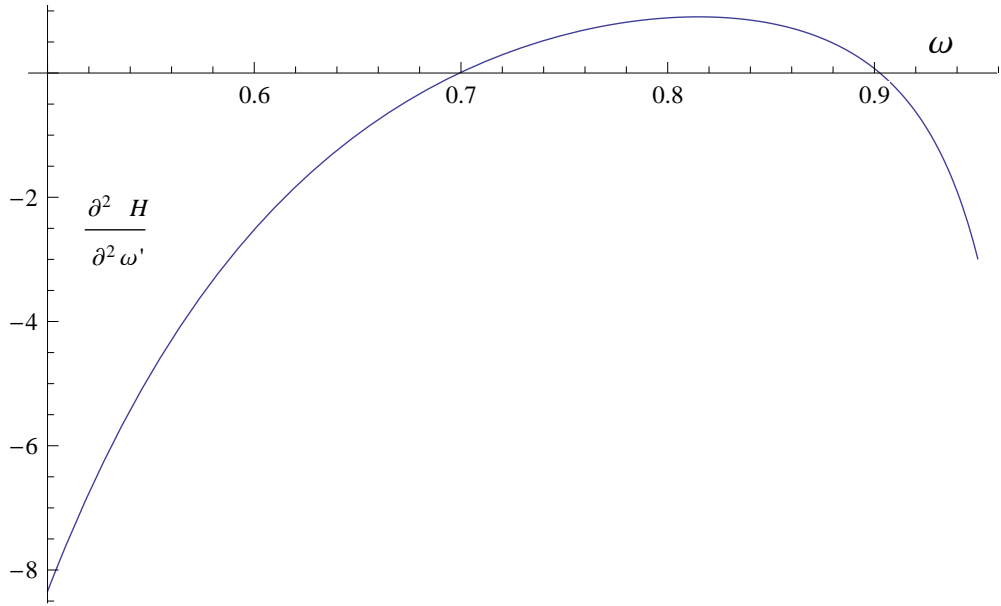}
    \caption{\label{5half}(Color online)
    Second derivative of probe Hamiltonian at $\omega'=\omega$ as a function of $\omega$ for $\kappa = 5/2$.}
    \end{figure}

 For $\kappa >2$ there is a second regime near the non-relativistic limit where the second derivative again becomes negative.
For example when $\kappa=5/2$ the second derivative  becomes negative both for $\omega <  0.699276 $ and in the non-relativistic regime $\omega > 0.902641.$  This is shown in Fig. \ref{5half}. This is in accord with the fact that for $\kappa>2$ the NLS  solutions are unstable to blowup.  However note that there is a range of $\omega$  where the second derivative is positive where  stability is not ruled out by this criterion.

For $ \kappa=1$ we have that 
\bq 
f(\omega',\omega) ^2 = \frac{\beta[\omega] \omega' }{\beta[\omega'] \omega} , 
\eq
where $\beta[\omega] = \sqrt{1-\omega^2}$.
The first derivative of $H_p$ with respect to $\omega'$ evaluated at $\omega' = \omega$ is  zero.
The second derivative evaluated at $\omega' = \omega$ leads to the following expression:
\bq
{H_p}_{\omega'\omega'}|_{\omega'=\omega} = -\frac{2 \left(\sqrt{1-\omega ^2} \left(\omega ^2-3\right)+4 \tanh
   ^{-1}\left(\sqrt{\frac{1-\omega }{\omega +1}}\right)\right)}{\omega ^2 \left(\omega
   ^2-1\right)^2}.
   \eq
   This function is zero at $\omega_B=0.697586$ and the second derivative is negative below this value of $\omega$.   (See Fig. \ref{one}). 
The values of $\omega_B$ vary  very slightly with $\kappa$. We find
 \ba
 &&\omega_B=0.703714 ~for~ \kappa=1/10  ;~~\omega_B =  0.699767 ~for~ \kappa=1/3;  \nonumber \\
&& \omega_B=0.698531 ~for~ \kappa=1/2  ;~~\omega_B =  0.697586 ~for~ \kappa=1;  \nonumber \\
&& \omega_B= 0.697963 ~for~ \kappa=3/2 ; ~~ \omega_B= 0.698612  ~for~ \kappa=2.
\ea
  One can view the probe Hamiltonian in a slightly different fashion.  Suppose we were choosing trial wave functions which have a fixed charge $Q=1$ in a time dependent variational approach to the problem.  Then we would choose as our trial wave functions to be 
  \bq
  \psi_v = \frac{\psi[\omega]}{\sqrt{Q[\omega,g^2]}}.
  \eq
Here $Q[\omega,g^2] = \int dx \psi^\dag \psi$. 
We would now find that the new Hamiltonian is given by 
\bq
H_v[\omega, g^2] = H_1[\omega] \left(\frac{1}{Q[\omega,g^2]} - \frac{1}{\kappa} \left(\frac{1}{Q[\omega,g^2]} \right)^{\kappa+1} \right) + \frac{ H_2[\omega] }{Q[\omega,g^2]}. 
\eq

Thinking now of $\omega$ as a variational parameter to be determined by the minimization of this Hamiltonian 
we would now {\it determine} $\omega$ as a function of $g^2$ by finding the stationary value of this Hamiltonian.

As an example let us choose $\kappa = 1$, where $\omega$  for fixed charge $Q$ is a function of $g^2$. Then 
\bq
H_v[\omega,g^2] =
-\frac{g^2 \omega  \left(\left(\omega ^2-1\right) \left(g^2 \omega -2 \sqrt{1-\omega
   ^2}\right)+2 \left(g^2 \omega  \sqrt{1-\omega ^2}+4 \omega ^2-4\right) \tanh
   ^{-1}\left(\sqrt{\frac{1-\omega }{\omega +1}}\right)\right)}{2 \left(1-\omega
   ^2\right)^{3/2}}. 
\eq
The first derivative is zero  when $g^2 [\omega]$ is given by Eq. (\ref{eq:Q1}), i.e.
      \bq
 g^2[\omega] = \frac{2 \sqrt{1- \omega^2}}{\omega}.
 \eq
 Also the second derivative of this Hamiltonian, evaluated at $ g^2[\omega]$   changes sign exactly at  $\omega_B=0.697586$.  This approach can be shown to be exactly equivalent to the Bogolubsky approach and yields the same values of $\omega_B$.

\subsection{Vakhitov-Kolokolov Criterion} \label{sec4.3}

\begin{figure}[ht!]
\begin{center}
\includegraphics[width=9.0cm]{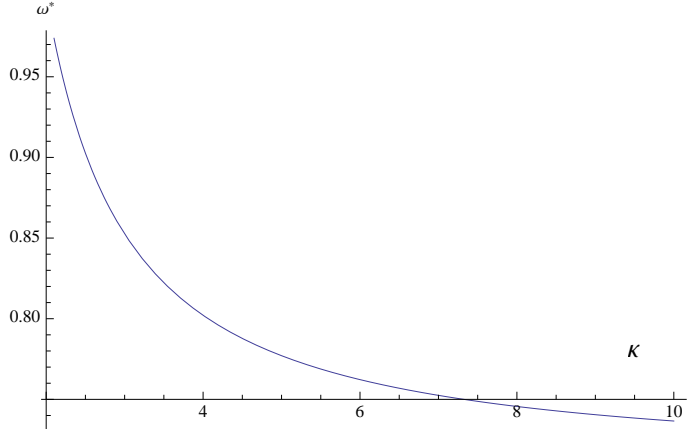} 
\end{center}
\caption{ $\omega^\star$ as a function of $\kappa$. For  $\omega \le \omega^\star $,  $dQ/d \omega < 0$ and there is no instability predicted for this deformation in this regime of $\omega$}
\label{omegastar} 
\end{figure}

In this section we will study the consequences of assuming that the Vakhitov-Kolokolov criterion, which was derived for the NLS equation, holds for the whole 
range of $\omega$ in the NLD case. That is we will explore the consequences of assuming 
one has stability when 
\bq
 \frac{dQ[\omega]}{d \omega} < 0, 
 \eq
 and instability otherwise.   For the NLD equation one has that 
 \ba
&& Q[\omega] =\frac{\sqrt{\pi } ((\kappa+1) (1-\omega))^{\frac{1}{\kappa}} \Gamma
   \left(1+\frac{1}{\kappa}\right)}{\kappa \omega (\omega+1) \sqrt{1-\omega^2}}   \nonumber \\
  && \times(\kappa+1) (\omega+1) \,
   _2{F}_1\left(-\frac{1}{2},1+\frac{1}{\kappa};\frac{3}{2}+\frac{1}{\kappa};
   \frac{1-\omega}{\omega+1}\right) \nonumber \\ 
   && + \omega (-\kappa+\omega-1) \,
   _2{F}_1\left(\frac{1}{2},1+\frac{1}{\kappa};\frac{3}{2}+\frac{1}{\kappa};\frac{1-\omega}{\omega+1}\right) ,  
   \ea
where $_2F_1$ is a hypergeometric function.  Taking the derivative, we find that for $\kappa <2$ it is always negative,
 suggesting that the solitary waves are stable in the entire range of $\omega$ values, 
 i.e. $0 < \omega < 1$. 
For $\kappa > 2$  one finds that there is a region of $\omega $ 
below the curve $\omega^\star(\kappa)$  where the solitary waves are suggested to be stable (Fig. \ref{omegastar}). However, both  suggestions will \textit{not} 
be confirmed by our simulations (Section \ref{sec5}). Thus the Vakhitov-Kolokolov criterion is not valid for the NLD case. 

 \section{Numerical methods} \label{sec5}

We have shown that different theoretical methods lead to 
different results on the stability of NLD solitary waves.
In order to understand and resolve these inconsistent results,
we  try to study numerically the stability of NLD solitary waves. 
We first tried a 4th order Runge-Kutta method which had worked very well for forced NLS equations with arbitrary nonlinearity exponent $\kappa$ \cite{cooper2012}. 
However, for the NLD equation we obtained inconsistent results, 
in particular for small values of $\kappa$. Various other numerical methods 
have been proposed in solving the NLD equation
and the readers are referred to a recent review \cite{XuShaoTang2013}.
It is also reported there that the operator splitting (OS) method
performs better than other numerical methods in terms of accuracy and efficiency.
The main advantage of the OS method is that 
different numerical techniques can be exploited into integrating the subproblems
in view of the features of the subproblems. 
In this work, we will employ the OS method to investigate the stability of NLD solitary waves.
The NLD system is decomposed into two subproblems, one is linear and the other one is nonlinear, 
and both of them can be integrated analytically with the non-reflection boundary condition (NRBC). For the sake of completeness, 
we will briefly describe below the OS scheme used in this paper, 
the related detailed theoretical analysis and numerical comparison with other schemes can 
be found in \cite{XuShaoTang2013}. 

For convenience, we rewrite the NLD system into
\begin{equation}
  \vPsi_t = \left(\mathcal{L} + \mathcal{N}\right)\vPsi,
\label{eq:split-4}
\end{equation}
where the linear operator $\mathcal{L}$ and  the nonlinear operator $\mathcal{N}$
are defined by
\begin{equation*}
{\mathcal{L}} \vPsi  := -\vgamma^0\vgamma^1 \vPsi_x,\quad
{\mathcal{N}} \vPsi :=  \mi \left(f-m\right)\vgamma^0\vPsi
\end{equation*}
with $f  := s (k+1) w^k$ and $w :=\wbar{\vPsi}\vPsi$.
In consequence, the problem \eqref{eq:split-4} may be decomposed into two subproblems as follows
\begin{align}
  \vPsi_t &= \mathcal{L} \vPsi,
  \label{eq:split-l-2}
  \\
  \label{eq:split-ns-2}
  \vPsi_t &= \mathcal{N} \vPsi.
\end{align}
Due to the local conservation law [see Eq.~\eqref{ns-conv} below] the 
solution of the nonlinear subproblem \eqref{eq:split-ns-2}
 may be expressed as an exponential of the operator $\mathcal{N}$
 acting on ``initial data''.
 Thus we may introduce the exponential operator splitting scheme
 for the NLD equation \eqref{eq:split-4},
 imitating that for the linear partial differential equations.
 Based on the exact or approximate solvers of those two subproblems, a more
general $K$-stage $N$-th order exponential operator splitting method
\cite{Sornborger1999}
for the system \eqref{eq:split-4} evolving from the $n$-th step to the $n+1$-th step 
can be cast into a product of  finitely many exponentials as follows
\begin{equation}
\label{eq:high-oder-split}
  \vPsi_j^{n+1} = \prod_{i=1}^{K}
  \big(
  \exp({\tau_{i} \mathcal{A}_i^{(1)}})\exp({\tau_{i}\mathcal{A}_i^{(2)}})
    \big) \vPsi_j^{n},
\end{equation}
where $\tau_i = a_i\tau$, with $\tau>0$ being the time stepsize,
denotes the time stepsize used within the $i$-th stage
and satisfies
$
\sum_{i=1}^K a_{i}=1,
$
and $\{\mathcal{A}_i^{(1)},\mathcal{A}_i^{(2)}\}$ is
any permutation of
$\{\mathcal{L},\mathcal{N}\}$.
The classical second-order Strang method \cite{Strang1968}
can be represented by $\widehat{\frac12}\widehat{\frac12}^T$
(\ie $a_i=\frac{1}{2}$ for $i=1,2$) 
if denoting $\widehat{a}_i:=\me^{\tau_i \mathcal{A}_i^{(1)}}\me^{\tau_i\mathcal{A}_i^{(2)}}$ and
$\widehat{a}_i^T:= \me^{\tau_i\mathcal{A}_i^{(2)}} \me^{\tau_i \mathcal{A}^{(1)}_i}$ \cite{Sornborger1999}. 
The remaining task is to determine the operators $\me^{\tau_i\mathcal{L}}$
and $\me^{\tau_i\mathcal{N}}$, \ie the solvers of the subproblems. 

The computational domain is set to be $[0, t_{fin}]\times[X_L, X_R]$.
Let $t_n = n\tau$ $(n=0,1,\ldots, t_{fin}/\tau)$ 
and $x_j = X_L + (j-1)h$ $(j = 1,2,\ldots,J)$ with $x_J=X_R$.
The ghost points are denoted by $x_0$ and $x_{J+1}$.
Here $\tau$ and $h$ are the time spacing and the spatial spacing, 
respectively.
\subsection{Linear subproblem}
\label{sec:dis-linear}

We now solve the linear subproblem \eqref{eq:split-l-2}.
We denote its ``initial data'' by $\vPsi_j^{(0)}=\big((\psi_1)_j^{(0)},(\psi_2)_j^{(0)}\big)^T$
at the $i$-th stage in \eqref{eq:high-oder-split} and its solution after $\tau_i$ by
$\vPsi_j^{(1)}=\big((\psi_1)_j^{(1)},(\psi_2)_j^{(1)}\big)^T$.
Denoting $\phi_1=\psi_1+\psi_2$ and $\phi_2=\psi_1-\psi_2$,
the linear subproblem \eqref{eq:split-l-2} can be rewritten as
\begin{equation} \label{eq:phi1phi2}
\begin{cases}
\partial_t \phi_1 + \partial_x \phi_1 = 0, \\
\partial_t \phi_2 - \partial_x \phi_2 = 0,
\end{cases}
\end{equation}
which means that the initial data of $\phi_1$ (resp. $\phi_2$)
simply propagate unchanged to the right (resp. left) with  velocity $1$.
Therefore (\ref{eq:phi1phi2}) can be \textit{exactly} integrated by 
the characteristics method with $\tau_i=h$ as follows
\begin{equation}
\begin{cases}
(\phi_{1})_j^{(1)} = (\phi_{1})_{j-1}^{(0)}, \\
(\phi_{2})_j^{(1)} = (\phi_{2})_{j+1}^{(0)},
\end{cases}
\end{equation}
with $j=1,\cdots, J$, 
and the values at the ghost points  
are naturally given by NRBC as
\begin{equation}\label{eq:nrbc}
\begin{cases}
(\phi_{1})_0 :=\phi_1(x_0,t)= 0, \\
(\phi_{2})_{J+1}: =\phi_2(x_{J+1},t) = 0,
\end{cases}
\end{equation}
where we have merely used the fact that 
outside a relatively big domain $[X_L,X_R]$,
the NLD spinor $\vPsi$ is negligibly small for 
it decays exponentially as $|x|\rightarrow +\infty$.
Consequently, we obtain the solution $\vPsi_j^{(1)}=\big((\psi_1)_j^{(1)},(\psi_2)_j^{(1)}\big)^T$ of the following form  
\begin{equation}
\begin{cases}
(\psi_{1})_j^{(1)} = \frac{(\phi_{1})_j^{(1)}+(\phi_{2})_j^{(1)}}{2}, \\
(\psi_{2})_j^{(1)} = \frac{(\phi_{1})_j^{(1)}-(\phi_{2})_j^{(1)}}{2}.
\end{cases}
\end{equation}

The characteristic method is very appropriate for the linear subproblem \eqref{eq:split-l-2}
only under the condition of $\frac{\tau_i}{h}$ to be an integer for all $i=1,\ldots,K$,
\ie all $a_i$ must be rational.
That is, the spatial spacing $h$ should be smaller than the time spacing $\tau$ which
results in huge computational cost. For example,
a fourth-order splitting with
rational $a_i$ demands $18$ stages given in \cite{Sornborger1999}
\begin{equation}\label{eq:4thsplit}
\widehat{\frac{1}{12}}^T\widehat{\frac{1}{12}}\widehat{\frac{1}{12}}^T\widehat{-\frac{1}{6}}
\widehat{\frac{1}{12}}^T\widehat{\frac{1}{12}}^T\widehat{\frac{1}{12}}^T\widehat{\frac{1}{12}}^T\widehat{\frac{1}{12}}
\widehat{\frac{1}{12}}^T\widehat{\frac{1}{12}}\widehat{\frac{1}{12}}\widehat{\frac{1}{12}}\widehat{\frac{1}{12}}
\widehat{-\frac{1}{6}}^T\widehat{\frac{1}{12}}\widehat{\frac{1}{12}}^T\widehat{\frac{1}{12}},
\end{equation}
and requires that $h=\frac{1}{12}\tau$,
which implies that  the number of grid points is $J=96000$
if choosing $\tau=0.025$ and $-X_L=X_R=100$.
To accelerate the simulations, 
we will adopt the multithread technology provided by OpenMP.
Note in passing that numerical results for the OS method are reported only 
for periodic boundary conditions with an irrational fourth-order splitting \cite{XuShaoTang2013}. 

\subsection{Nonlinear subproblem}

The nonlinear subproblem \eqref{eq:split-ns-2} is
left to be solved now.
Its ``initial data'' is still denoted by $\vec\Psi_j^{(0)}=\big((\psi_1)_j^{(0)},(\psi_2)_j^{(0)}\big)^T$
at the $i$-th stage in \eqref{eq:high-oder-split}, and define
$$
t_n^{(i)}=t_n+\sum_{p=1}^{i-1} \tau_p,\quad i=1,2,\cdots,K.
$$ 
For the nonlinear subproblem \eqref{eq:split-ns-2},   
it is not difficult to verify that
\begin{equation}\label{ns-conv}
\partial_t w = 0, \quad \partial_t f = 0.
\end{equation}
Using this local conservation law gives \textit{analytically} the solution at $t=t_n^{(i+1)}$ of \eqref{eq:split-ns-2}
with the ``initial data'' $\vec\Psi_j^{(0)}$ as follows
\begin{align}
\vPsi_j^{(1)}
&= \exp\left(\mi \int_{t_n^{(i)}}^{t_n^{(i+1)}} (f-m)_j\vgamma^0 \dif t \right) \vPsi_j^{(0)}
= \exp\left(\mi (f-m)_j^{(0)}\vgamma^0 \tau_{i} \right) \vPsi_j^{(0)} \nonumber \\
&= \diag\left\{\exp\left( \mi (f-m)_j^{(0)} \tau_{i}\right),
\exp\left(-\mi(f-m)_j^{(0)} \tau_{i}\right)\right\} \vPsi_j^{(0)}. \label{OS-NS-eq3}
\end{align}

With NRBC, 
subproblems \eqref{eq:split-l-2} and \eqref{eq:split-ns-2} can be both solved analytically and the numerical error only comes from the operator splitting in time. 
That is, the OS method with the rational splitting \eqref{eq:4thsplit} (recall that the spatial spacing $h=\frac{\tau}{12}$), 
denoted by OS(4) hereafter, 
is of the order $\mathcal{O}(\tau^4)$,
which is confirmed numerically by simulating a normalized standing wave with $\kappa=1$, $\omega=0.50$ and the centroid located at $x=0$,  
see Columns 2-5 of Table~\ref{tab:accuracycheck},
where $\text{err}_2$ and $\text{err}_\infty$ are 
the $l^2$ and $l^\infty$ errors, respectively. 
The centroid position $q(t)$ does not change at all until $t=100$,
see Column 6 of Table~\ref{tab:accuracycheck}.
We have also shown there that 
$\mathcal{V}_Q$, $\mathcal{V}_E$, $\mathcal{V}_P$,
measuring  respectively
the variation of charge, energy and linear momentum
at $t=100$ relative to the initial quantities, 
are all almost zero,
see Columns 7-9, which demonstrates that  
the OS(4) method is able to keep the charge, energy and linear momentum
constant before the instability happens.  
(In fact, it will be shown later that 
this normalized standing wave is unstable and the instability appears at $t=11036$, 
see Fig.~\ref{fig:k1_omega0.5}). 
We can conclude that the OS(4) method is highly accurate and
the numerical error is controlled only by the time step size $\tau$
for no approximation is used in space.

To perform the numerical study of the stability of NLD solitary waves,
the employed numerical method is required to be not only of high-order accuracy
but also immune to the effect of artificial boundaries $X_{L,R}$. 
NRBC~\eqref{eq:nrbc} used in the OS(4) method
can avoid completely 
the numerical effect of $X_{L,R}$
on the stability of NLD solitary waves provided 
a relatively big domain $[X_L,X_R]$ is adopted,
since it is transparent for outgoing waves 
and does not allow any waves to be pumped into the computational domain. 
In such situations, we can also prove easily that the OS(4) method conserves the total charge. 
In summary, 
the proposed OS(4) method with NRBC 
is  very appropriate and will be used for investigating the stability of NLD solitary waves. 

\begin{table}
\centering
\caption{\small Accuracy check for the OS method with NRBC and a rational fourth-order splitting. We take a normalized solitary wave with $\kappa=1$ and $\omega=0.50$ as an example, 
and measure the related quantities within the domain $[-100,100]$ at $t=100$.
Here  
$\tau$ is the time stepsize, 
$\text{err}_2$ and $\text{err}_\infty$ are respectively
the $l^2$ and $l^\infty$ errors,
$q$ denotes the centroid position of charge density,
$\mathcal{V}_Q$, $\mathcal{V}_E$, $\mathcal{V}_P$
measure respectively the variation of charge, energy and linear momentum
at the final time relative to the initial quantities.
}
\begin{tabular}{|c|c|c|c|c|c|c|c|c|}
\hline
$\tau$   & $\text{err}_{2}$  & Order &  $\text{err}_{\infty}$ & Order &  $q$ & $\mathcal{V}_{Q}$ &       $\mathcal{V}_{E}$ &       $\mathcal{V}_{P}$\\
\hline
0.1    & 2.99E-09 &        & 2.12E-09 &       & 2.75E-14 & 2.22E-16 & 2.22E-16 & 2.28E-16 \\
\hline
0.05   & 1.86E-10 & 4.01   & 1.32E-10 & 4.01  & 3.20E-15 & 1.78E-14 & 8.33E-15 & 3.57E-17 \\
\hline
0.025  & 1.16E-11 & 4.00   & 8.24E-12 & 4.00  & 1.33E-14 & 7.66E-15 & 3.44E-15 & 2.30E-16 \\
\hline
0.0125 & 7.26E-13 & 4.00   & 5.87E-13 & 3.81  & 1.23E-14 & 1.14E-13 & 5.55E-14 & 2.98E-16 \\
\hline
\end{tabular}\label{tab:accuracycheck}
\end{table}

\section{Numerical results}

\begin{figure}
\centering
\subfigure[$\kappa=0.1$]{\includegraphics[width=6cm]{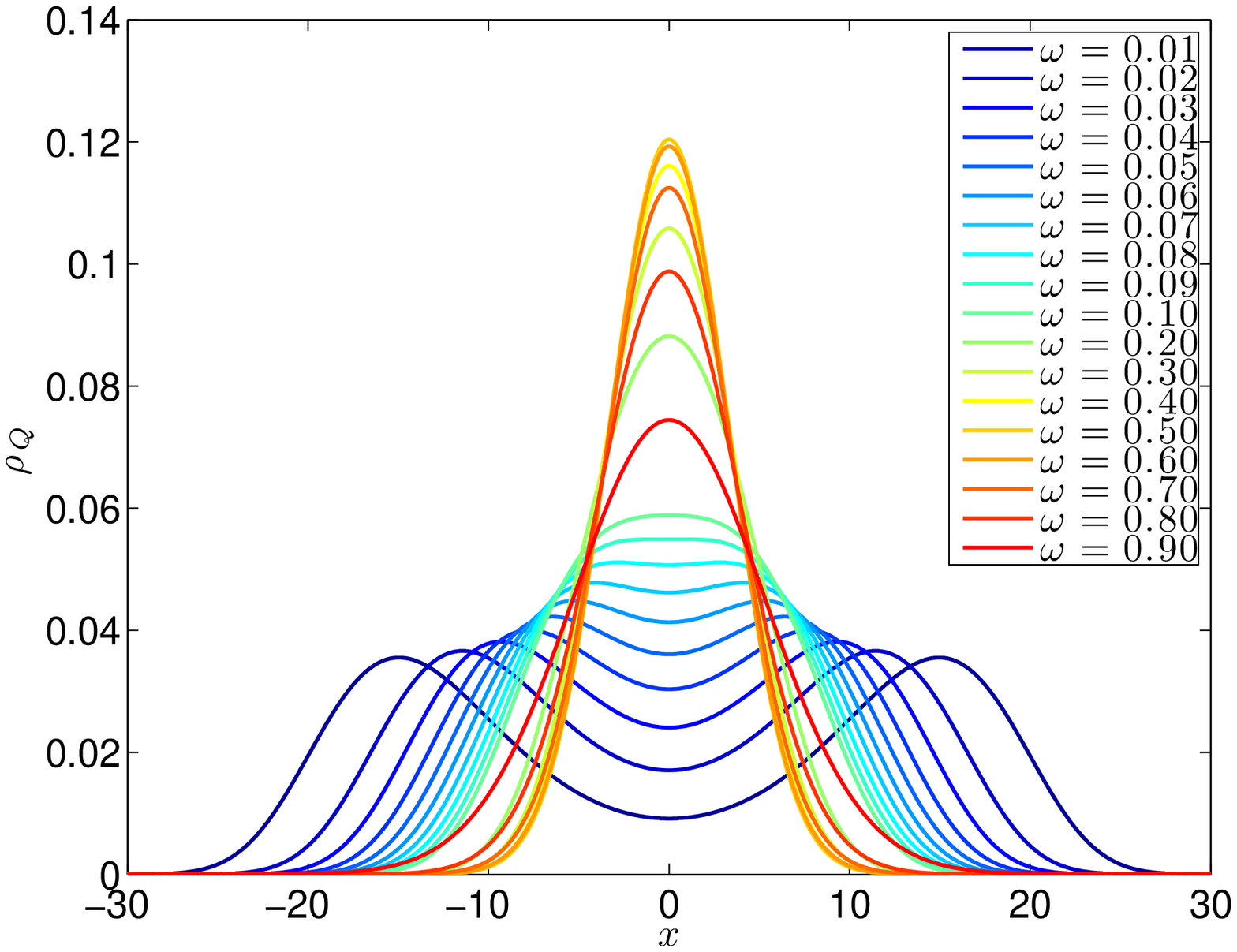}}
\subfigure[$\kappa=0.5$]{\includegraphics[width=6cm]{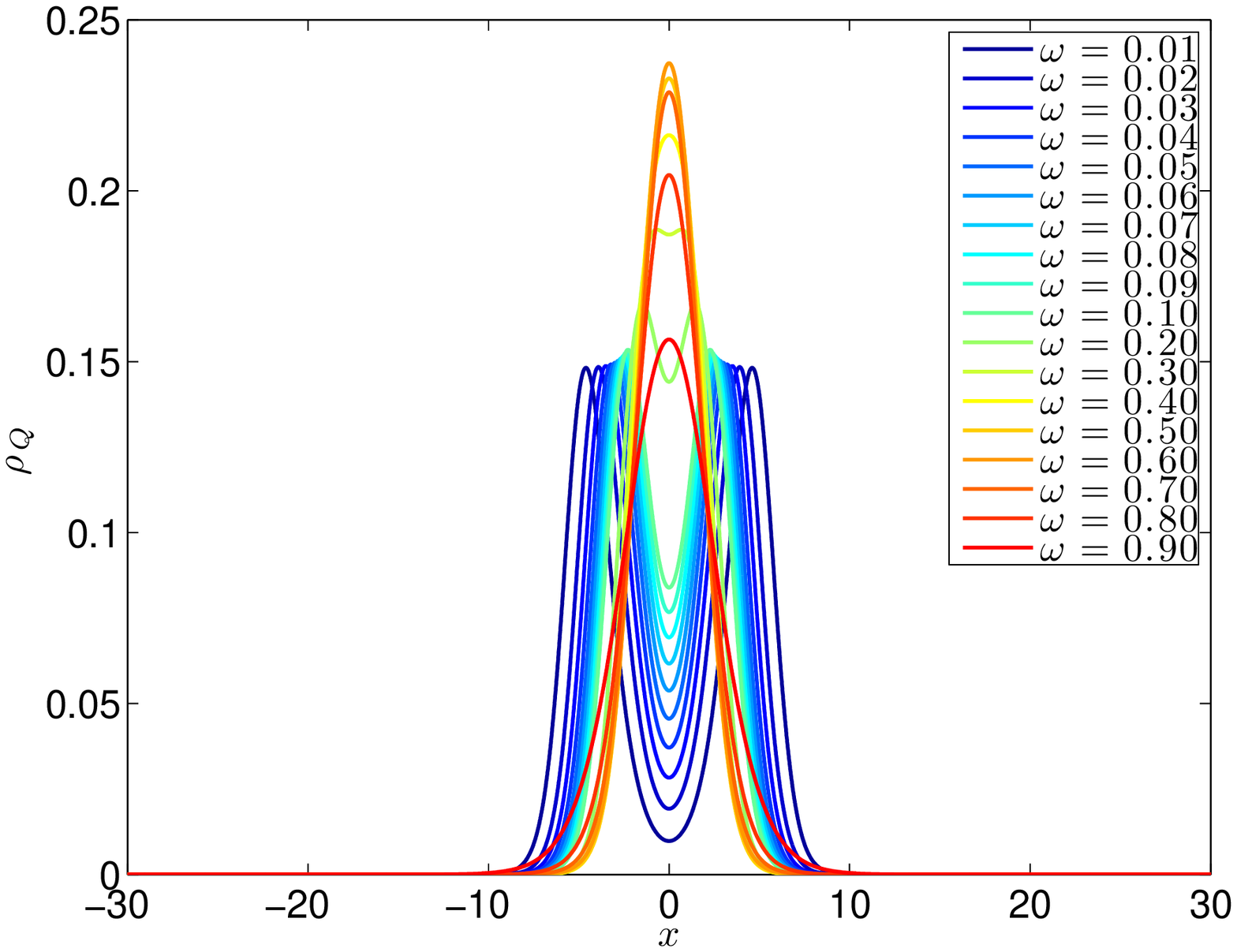}}\\
\subfigure[$\kappa=1.0$]{\includegraphics[width=6cm]{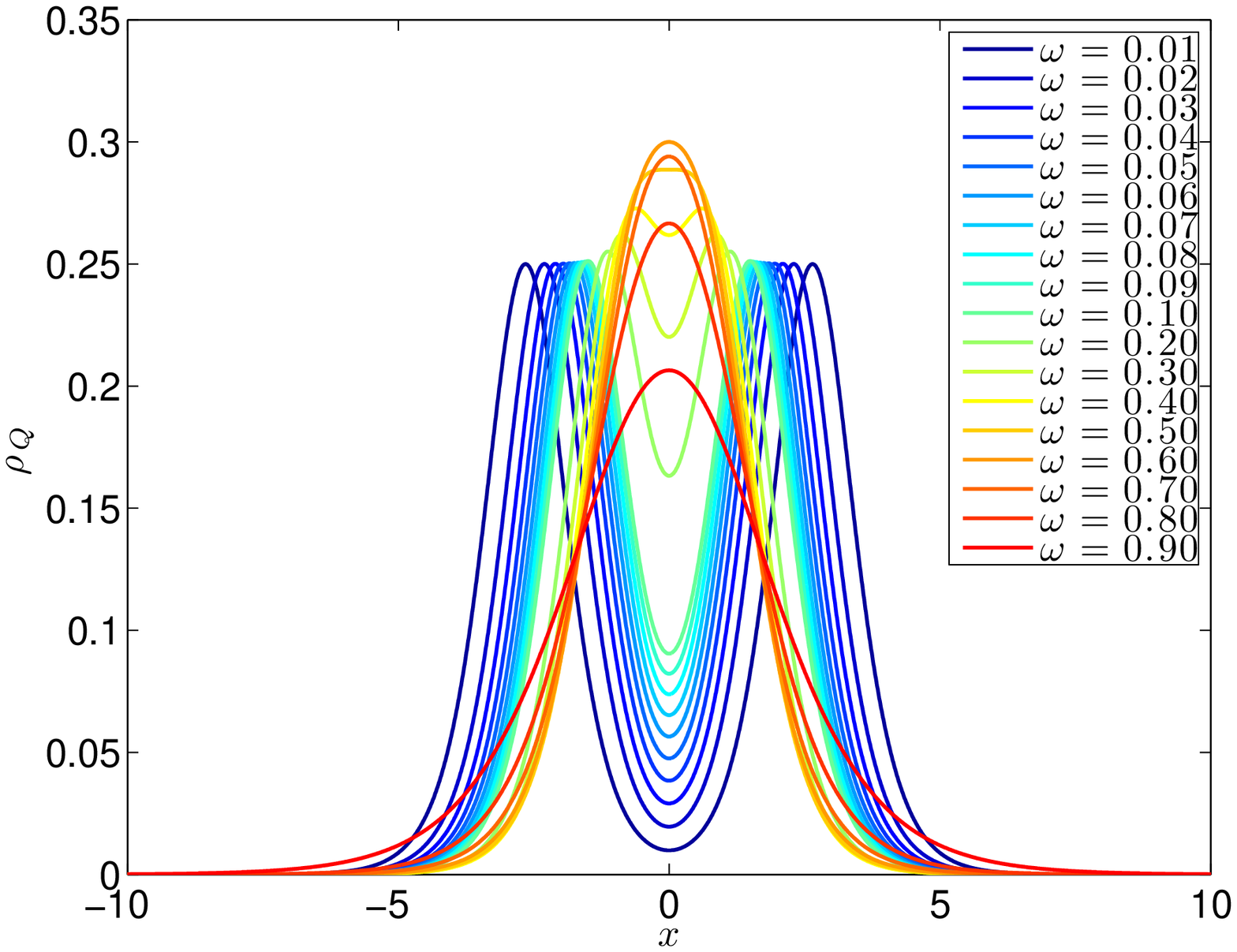}}
\subfigure[$\kappa=1.5$]{\includegraphics[width=6cm]{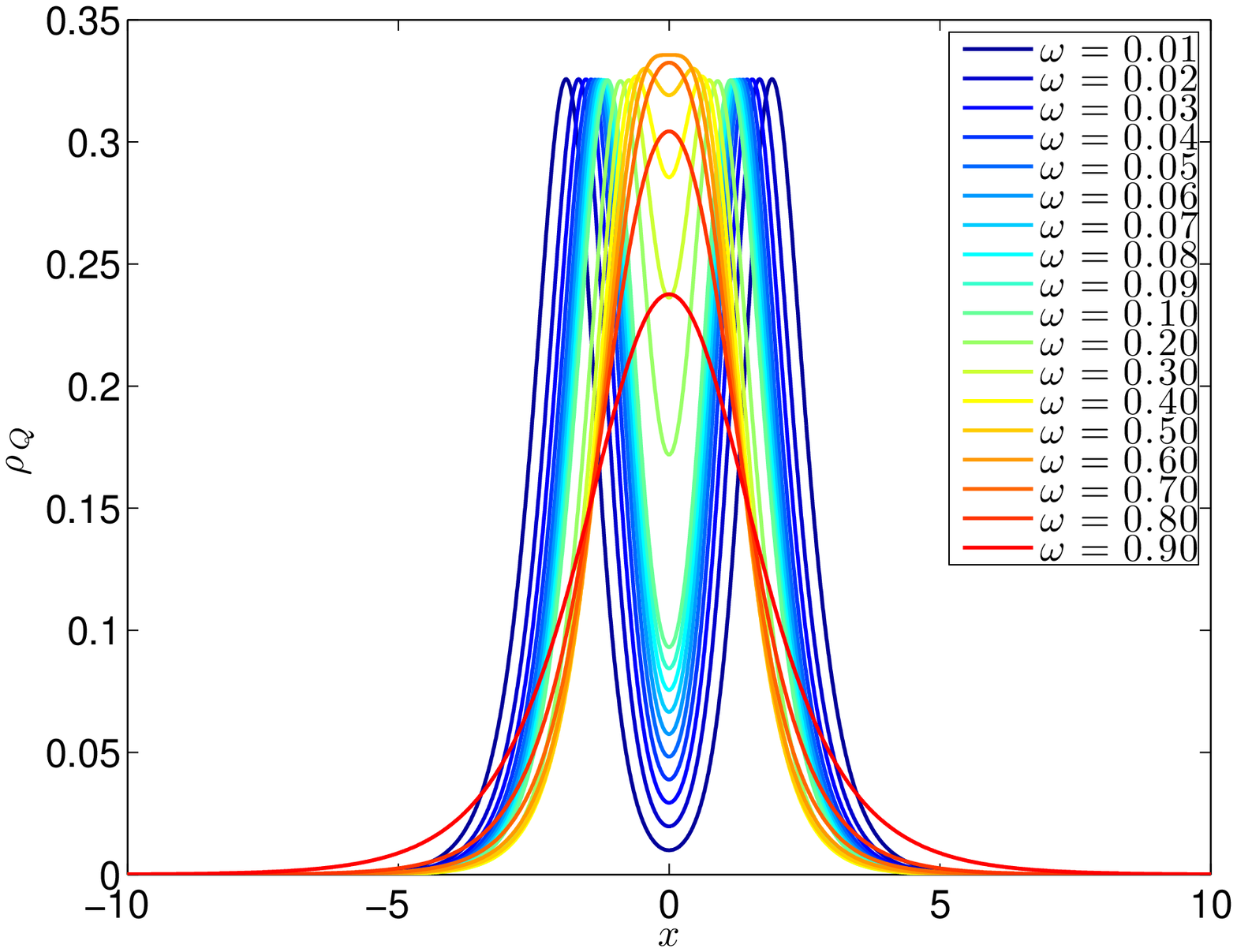}}\\
\subfigure[$\kappa=2.0$]{\includegraphics[width=6cm]{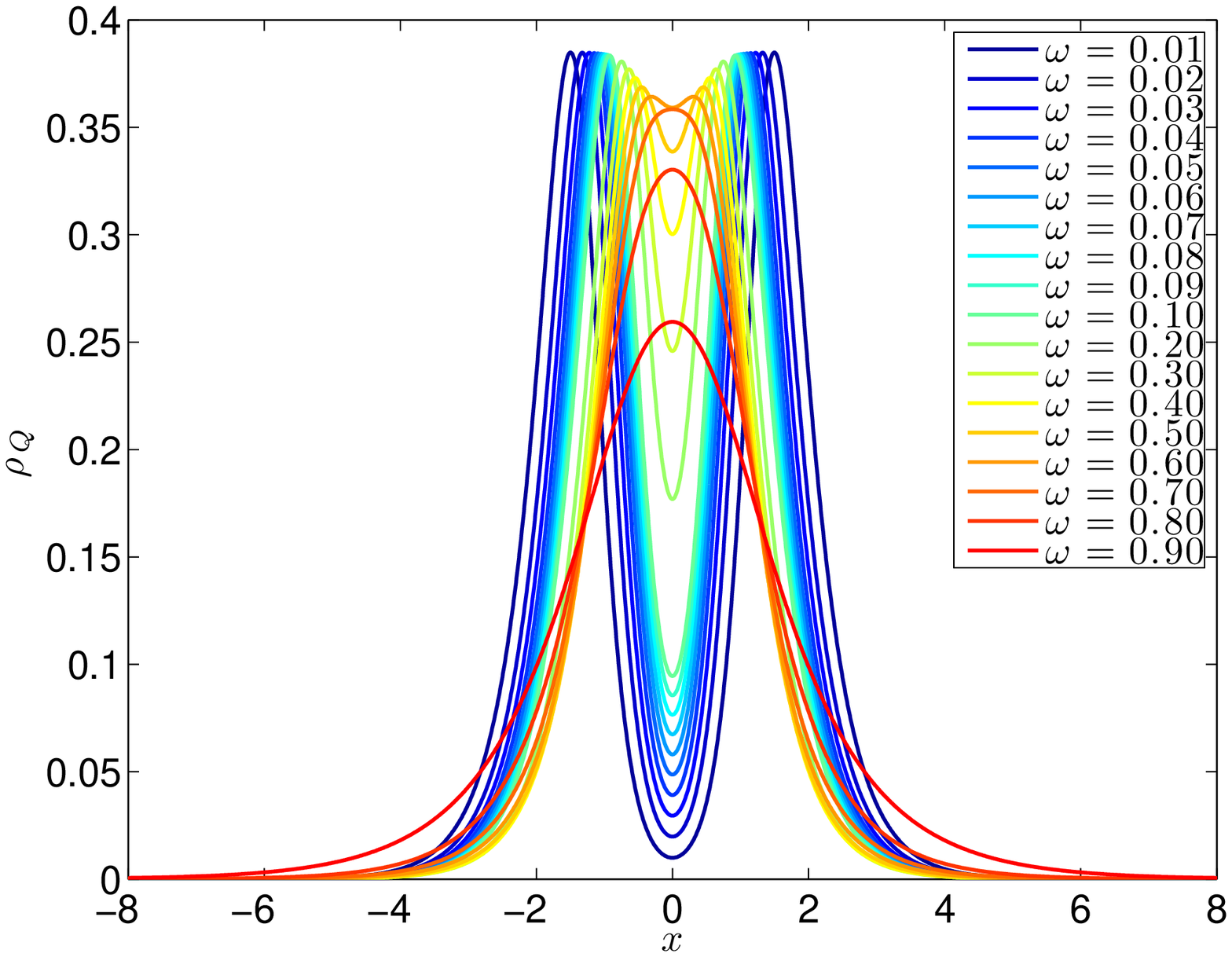}}
\subfigure[$\kappa=2.4$]{\includegraphics[width=6cm]{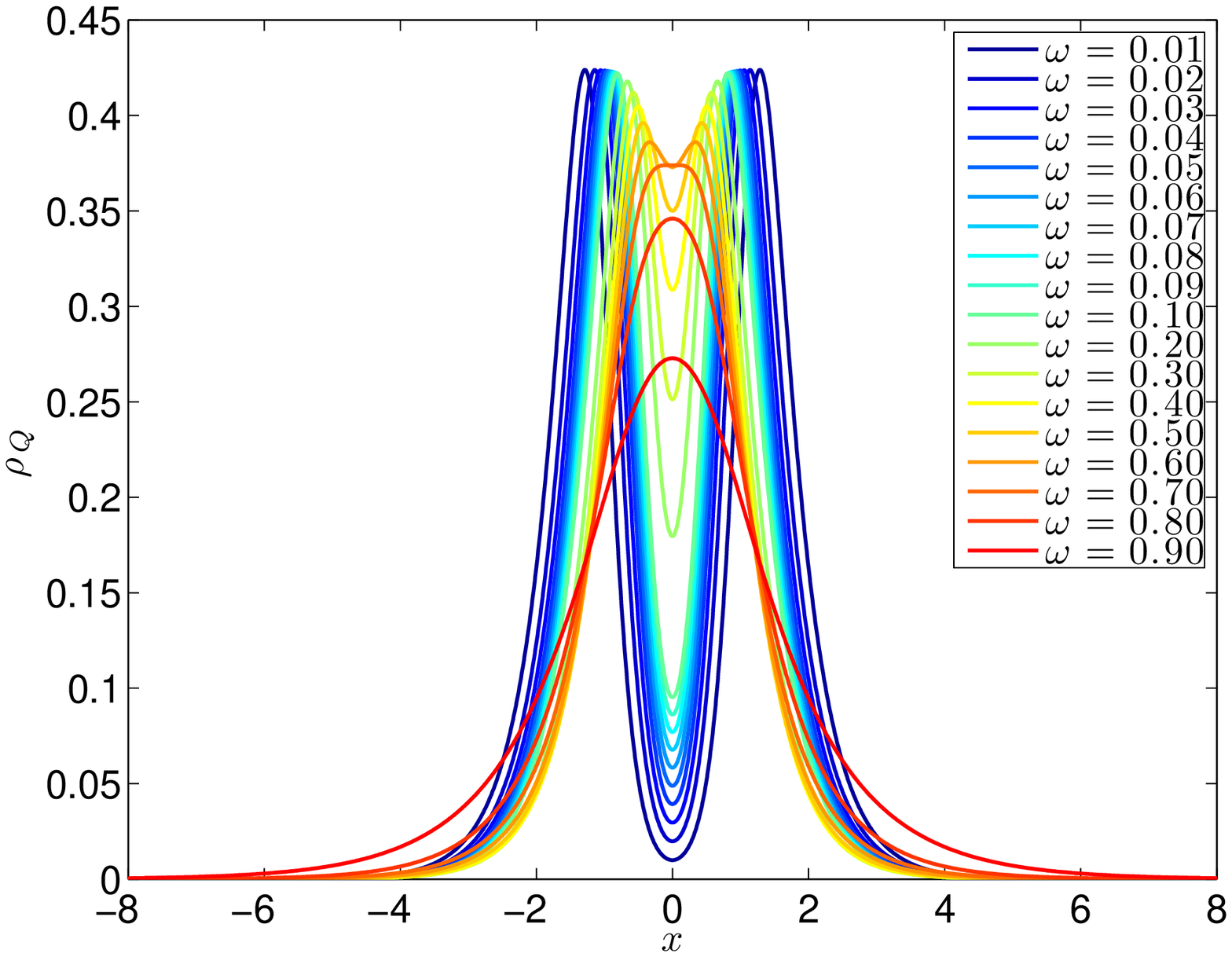}}
\caption{\small Typical profiles of the charge density $\rho_Q$ for various exponent powers (or the nonlinearity parameter)  $\kappa$ 
and frequencies $\omega$.}
\label{fig:profile}
\end{figure}

In accordance with the theoretical results, 
we consider merely the normalized NLD solitary waves, 
\ie the charge is fixed to be $Q\equiv 1$.
For such normalized NLD waves, 
only the frequency $\omega$ can be adjusted to get different profiles 
if fixing the mass $m=1$ and the exponent power (or the nonlinearity parameter) $\kappa$. 
For $\kappa=0.1,0.5,1.0,1.5,2.0,2.4$, Fig.~\ref{fig:profile}
plots the profile transition of charge density $\rho_Q$
when $\omega$ increases from $0.01$ to $0.9$.
It is clearly observed there that,
as the frequency increases,
the charge density is transmitted from a two-hump profile to a one-hump profile
during which the valley of the two-hump wave rises 
until the one-hump wave is formed and then disappears;
the maximum height of the peak of the one-hump wave is larger than that of the two-hump wave for $\kappa=0.1,0.5,1.0$, comparable for $\kappa=1.5$ and less than for $\kappa=2.0,2.5$.
Actually, it has been proved that the charge density has either one hump or two humps under 
the pure scalar self-interaction and also conjectured that 
there is a connection between the stability and the multi-hump structure \cite{NLDE,XuShaoTangWei2013}. In the following 
we will use the OS(4) method with NRBC to study 
such stability of normalized NLD waves and 
determine the range of $\omega$ in which 
the NLD solitary waves are stable or unstable for a given nonlinearity (or exponent power) $\kappa$.
For simplicity, we only consider here the standing waves with the centroid located at $x=0$. 

\subsection{$\kappa=1$}

\begin{figure}
\centering
{\includegraphics[width=6cm]{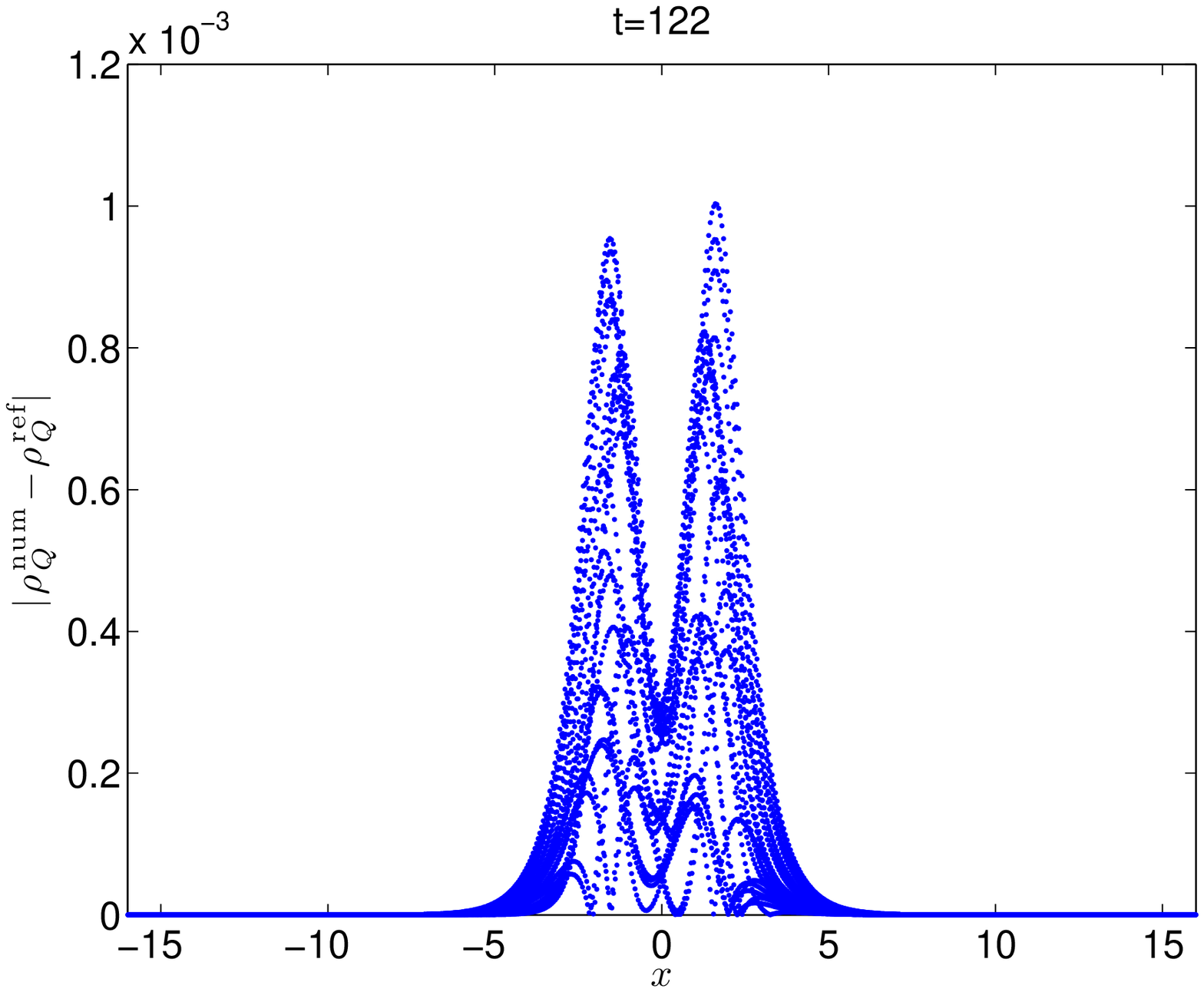}}
{\includegraphics[width=6cm]{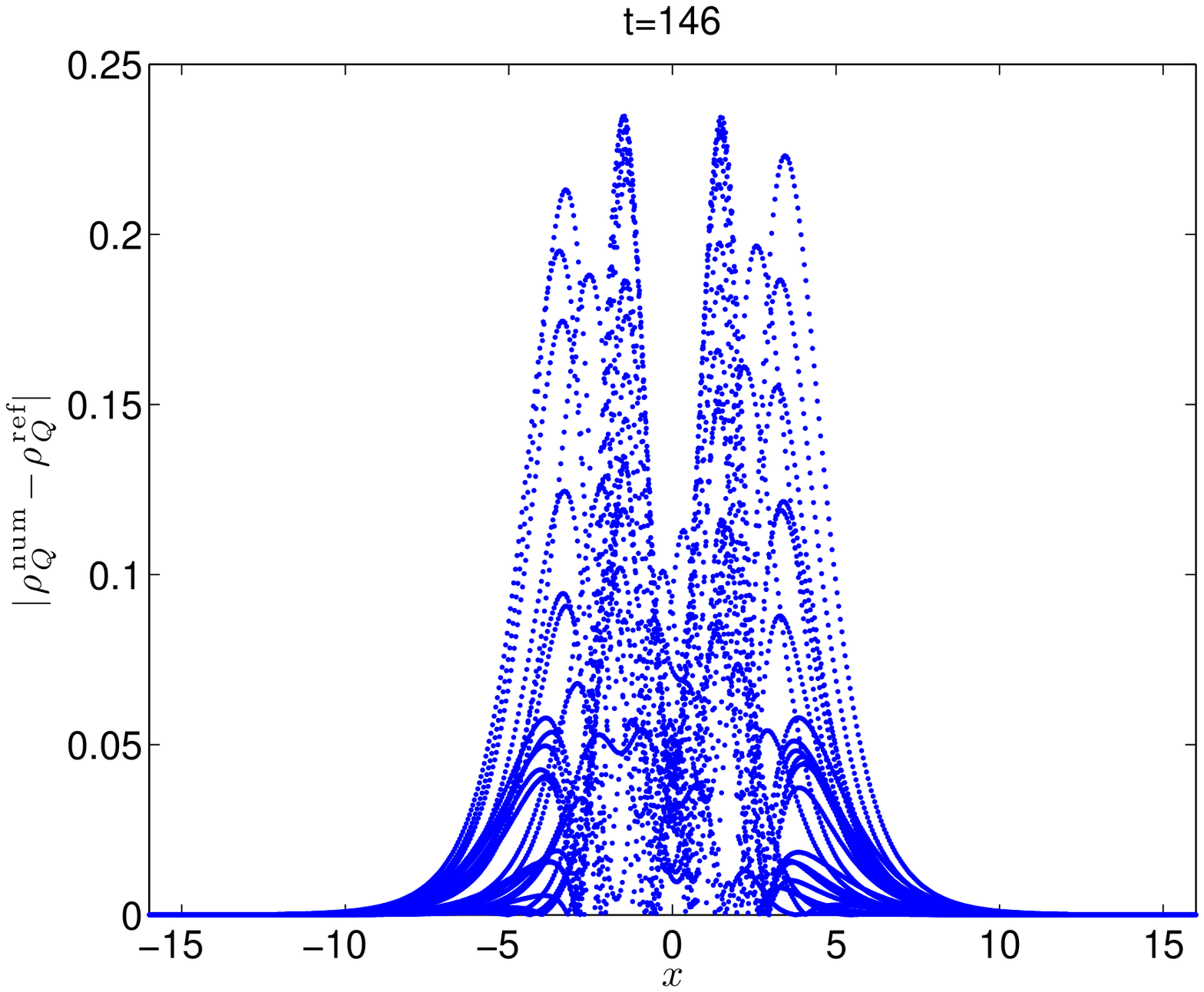}}
\caption{\small Unstable two-hump solitary wave with $\kappa=1$ and $\omega=0.1$: Snapshot of the difference of the charge density between the numerical solution and the reference solution at $t=t_e=122$ and $t=t_c=146$, (left and right panels,  respectively). Here $\tau=0.025$ and $L=100$. 
}
\label{fig:k1_omega0.1}
\end{figure}

\begin{figure}
\centering
\subfigure[$\omega=0.1$]{\includegraphics[width=6cm]{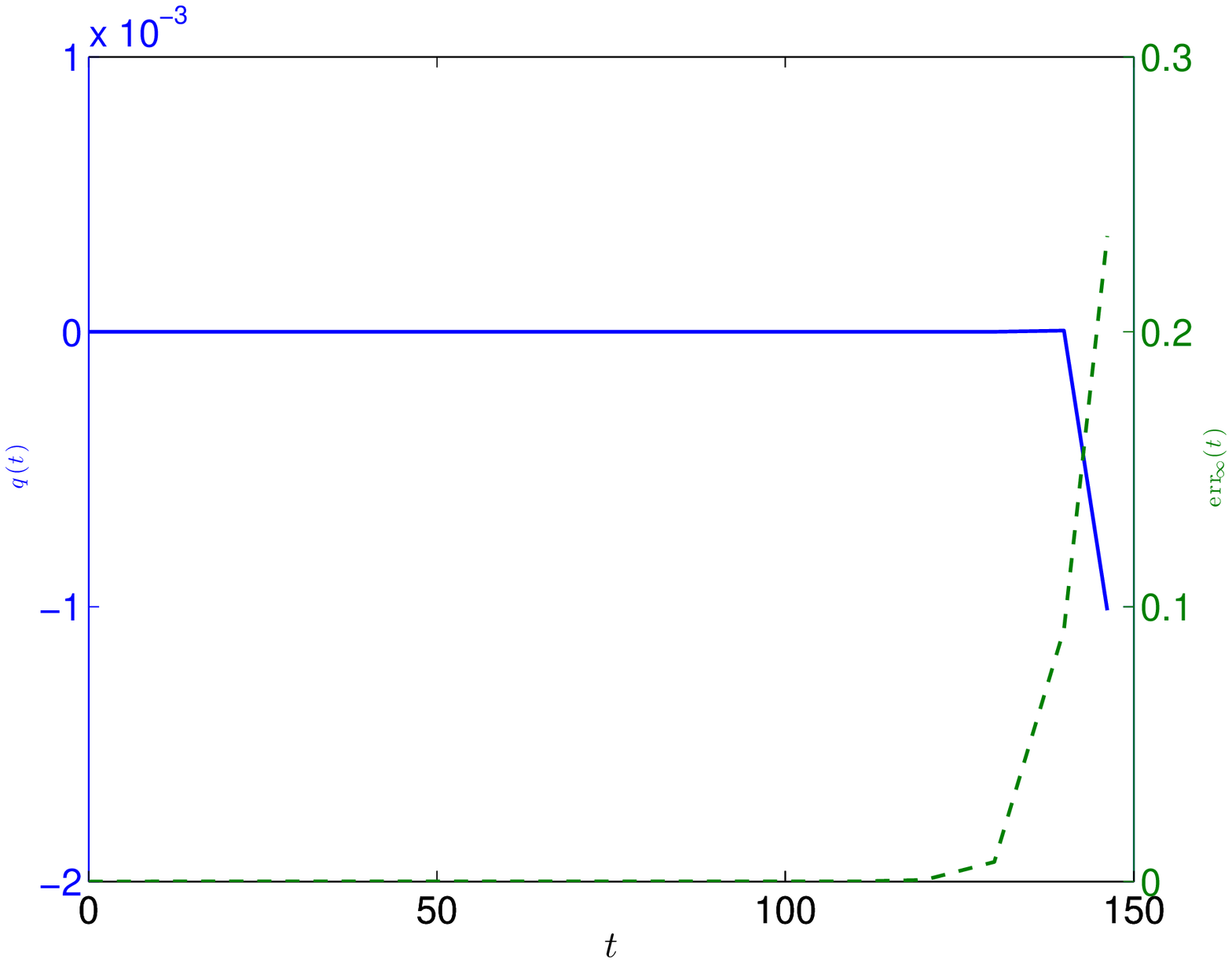}}
\subfigure[$\omega=0.5$]{\includegraphics[width=6cm]{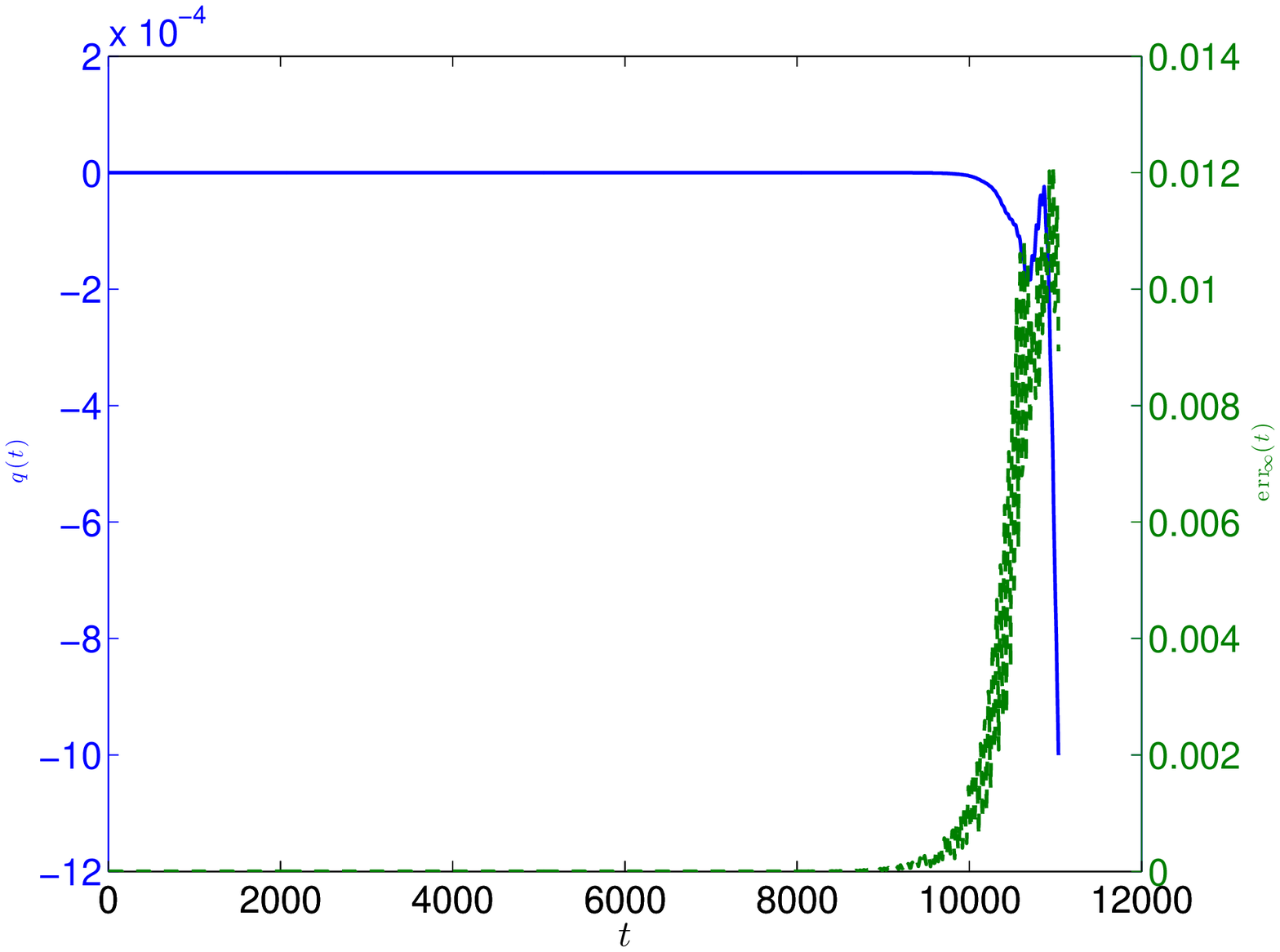}}
\caption{\small Plots of the centroid position $q(t)$ (solid line) and 
the $l^\infty$ error $\text{err}_\infty$ (dashed line) vs.  time for $\kappa=1$. Here $\tau=0.025$ and $L=100$.}
\label{fig:k1_error}
\end{figure}

In this section, we present the numerical results for the Soler model \cite{alvarez} \ie $\kappa=1$. The first numerical simulation is performed using the OS(4) method for the two-hump wave with $\omega=0.1$, 
see Fig.~\ref{fig:profile}(c). 
A large computational domain $[-L,L]$ (\ie $X_L=-L,X_R=L$) is set with
$L=100$. The time spacing, the parameter controlling the numerical error,
is taken to be $\tau=0.025$. That is,  the numerical error 
introduced by the OS(4) method at each time step is about $\tau^4\simeq 3.91$E-$07$. However, the numerical error often accumulates slowly over time.
If the solitary wave is unstable, such a slowly accumulated numerical error will 
be amplified in a relatively short period, after that the wave will change its position which implies that the instability happens.   
This indeed occurs when $\omega=0.1$, see Row 5 of Table~\ref{tab:Ldependent}.
There we have shown the instants of time at which the monitored quantities, 
$q$, $\text{err}_{\infty}$, 
$\text{err}_{2}$, 
$\mathcal{V}_{P}$,       
$\mathcal{V}_{E}$,  
become larger than a given tolerance $\epsilon$ ($=1.00$E-$03$ here).
It can be seen that $\text{err}_{\infty}$, $\text{err}_{2}$, 
$\mathcal{V}_{E}$, $\mathcal{V}_{P}$, and $q$ increase over 
$\epsilon$ in sequence. We denote the instant at which the centroid position $q$ (resp. $\text{err}_{\infty}$)
becomes larger than $\epsilon$ by $t_c$ (resp. $t_e$).
In Fig.~\ref{fig:k1_omega0.1},   
we plot the difference of the charge density between the numerical solution and the reference solution at $t_e=122$ and $t_f=146$, respectively. 
Meanwhile, the history of 
$q$ and $\text{err}_{\infty}$ is displayed in Fig.~\ref{fig:k1_error}(a). 
It is observed there that, although the accumulated numerical error is larger than $\epsilon$ at $t_e=122$, the NLD wave still preserves its two-hump shape and its centroid hardly wavers from the initial position; after that,
$\text{err}_{\infty}$ increases quickly, soon the wave loses its shape,
many waves are then generated and 
the centroid moves from $x=0$ over $\epsilon$ at $t_f=146$.
Hereafter, we define $t_c$ to be the moment at which the instability sets in.
As shown clearly in Fig.~\ref{fig:k1_omega0.1},
the entire process from $\text{err}_{\infty}>\epsilon$ to $q>\epsilon$
develops very fast because it takes place only in the central area (around the initial centroid position $x=0$). This is also confirmed by numerical simulations within the domain $[-L,L]$ of different length, say $L=50,75,100,125,150$, 
which reveal that 
instants of time at which the monitored quantities 
become larger than $\epsilon$ are nearly independent of the domain length,
see Rows 2-7 of Table~\ref{tab:Ldependent}. 
During the process, no charge is radiated out from the central area and thus the total charge is conserved, \eg at $t_f=146$ 
$\mathcal{V}_{Q}\simeq 2.23$E-$14$ for $L=75$ and 
$\mathcal{V}_{Q}\simeq 9.66$E-$15$ for $L=100$.

\begin{figure}
\centering
{\includegraphics[width=6cm]{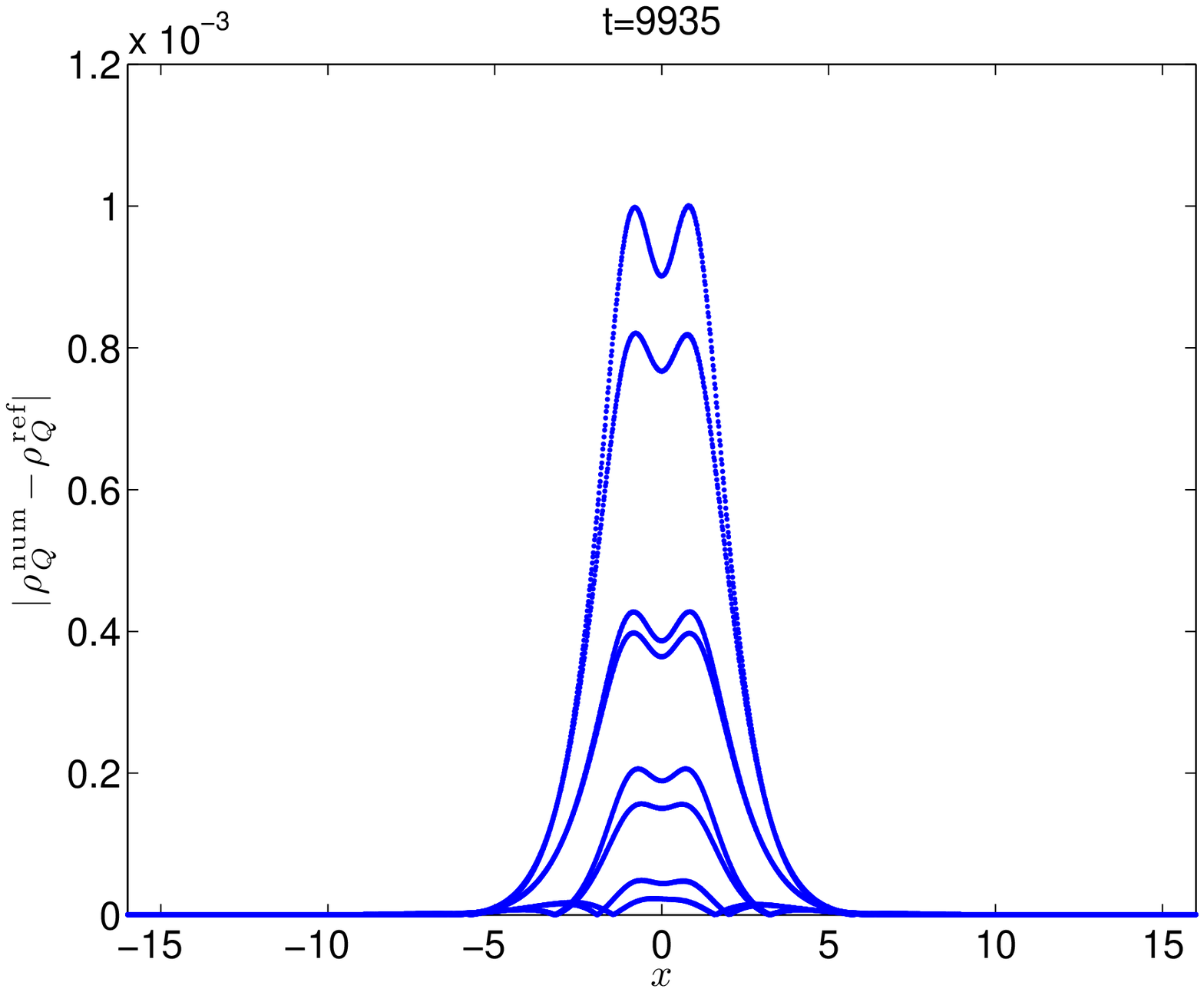}}
{\includegraphics[width=6cm]{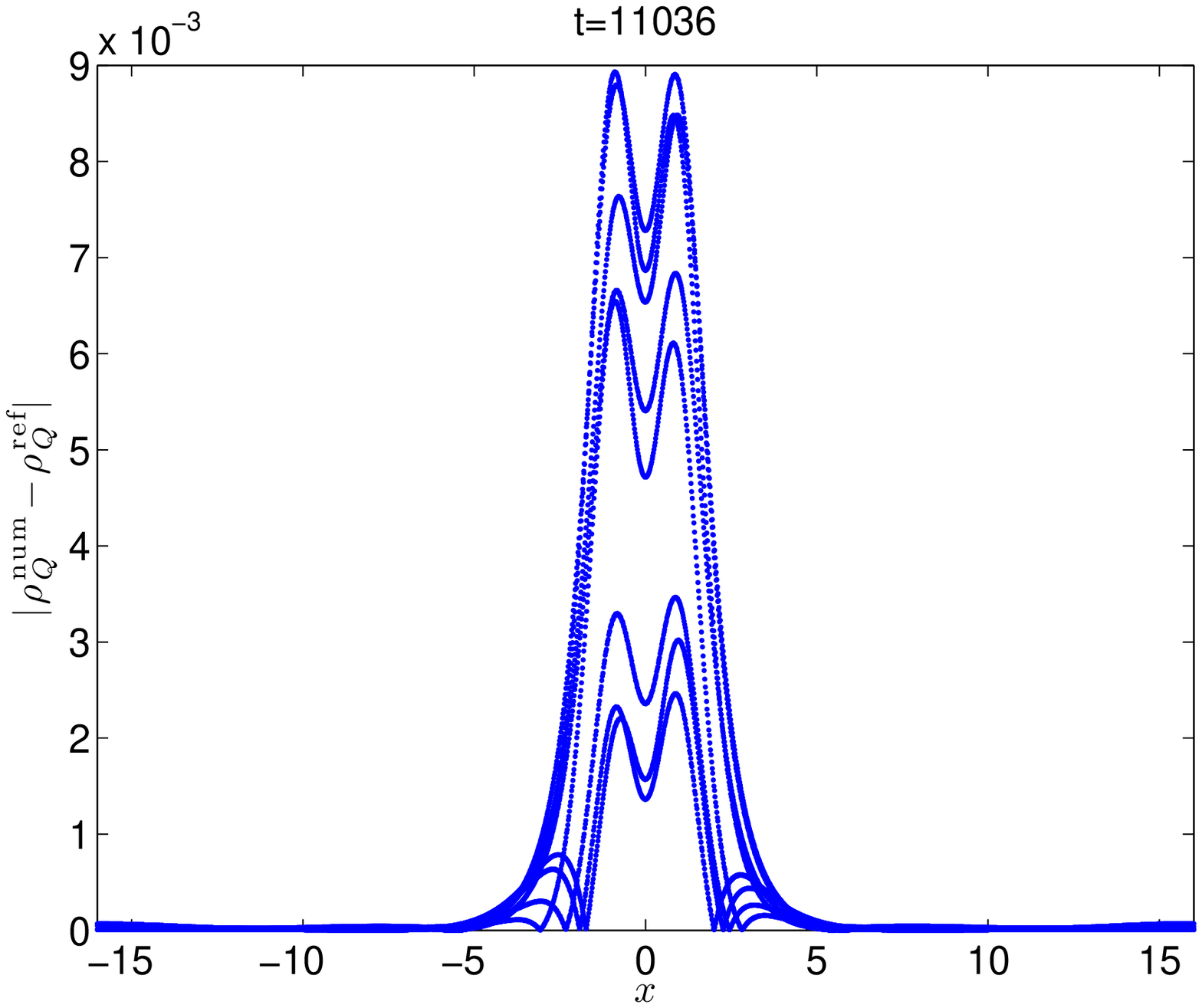}}
\caption{\small Unstable one-hump solitary wave with $\kappa=1$ and $\omega=0.5$: Snapshot of the difference of the charge density between the numerical solution and the reference solution at $t=t_e=9935$ and $t=t_c=11036$,  (left and right panels,  respectively). Here $\tau=0.025$ and $L=100$.}
\label{fig:k1_omega0.5}
\end{figure}

\begin{figure}
\centering
\includegraphics[width=8cm]{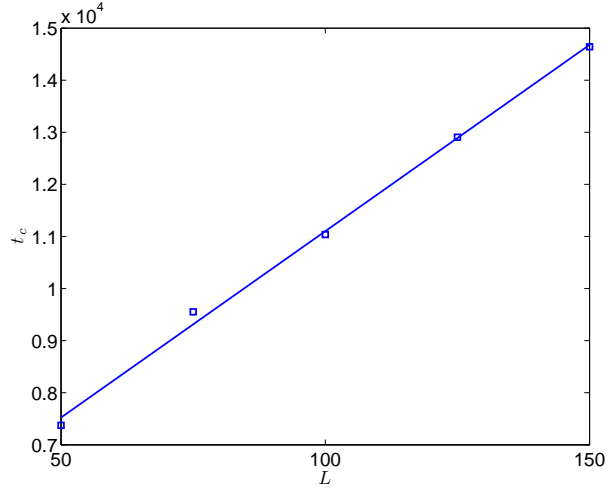}
\caption{\small Plot of $t_c$ at which the centroid position $q(t)$ becomes larger than $\epsilon$ ($=1.00$E-$03$ here) with respect to $L$ for $\kappa=1$ and $\omega=0.5$. 
The computational domain is $[-L,L]$ and five different lengths are tested.
The concrete data are given in Table~\ref{tab:Ldependent}. 
Here $\tau=0.025$.}
\label{fig:k1_L}
\end{figure}

The second numerical simulation is performed 
for the one-hump wave with $\omega=0.5$, see Fig.~\ref{fig:profile}(c). 
The setup of the OS(4) method for simulating the two-hump wave with 
$\omega=0.1$ is used. We plot the dfference of the charge density between the numerical solution and the reference solution at $t_e=9935$ and $t_f=11036$ 
in Fig.~\ref{fig:k1_omega0.5}  
as well as the history of $q$ and $\text{err}_{\infty}$ in Fig.~\ref{fig:k1_error}(b).
Therefore this one-hump wave is considered to be unstable.
However, contrary to the fast process occuring only in the central area 
when $\omega=0.1$,
the entire process from $\text{err}_{\infty}>\epsilon$ to $q>\epsilon$
develops very slowly when $\omega=0.5$.
As demonstrated by Figs.~\ref{fig:k1_omega0.5} and \ref{fig:k1_error}(b),
the reason for such a slow process is the following:   
Although many waves of small amplitude are generated because of the instability, the 
wave is  unstable  
only if enough generated waves move outside the computational domain. 
This is further confirmed by numerical simulations within  domains of different 
lengths the results of which can be found in 
Rows 8-13 of Table~\ref{tab:Ldependent}.
Those results show that
the variation of charge $\mathcal{V}_Q$ decreases by $\epsilon$
before the instability occurs at $t_c$; and $t_c$ linearly depends on $L$ as plotted in Fig.~\ref{fig:k1_L}. 

We have shown above that the OS(4) method with NRBC 
is capable of capturing the instability regardless of whether it occurs quickly or slowly.
When the time step size $\tau$, the only parameter controlling the numerical error, decreases from $0.025$ to $0.0125$, 
we have a very small change of $t_c$, \eg 
$t_c=146$ (resp. $t_c=11306$) for $\tau=0.025$ and $t_c=148$ (resp. $t_c=11278$) for $\tau=0.0125$ when $\omega=0.10$ (resp. $\omega=0.50$).
Consequently, 
the methodology to determine the stable range for $\omega$ [denoted by $\Omega_\kappa$, being a subset of $(0,1)$] in which 
the NLD waves are stable, is to use the OS(4) method with $\tau=0.025$ and $L=100$ to
simulate the wave with the frequency $\omega_0$. If the centroid position $q(t)$ is always less than the given tolerance $\epsilon$ before a prescribed final time $t_{fin}$,
then $\omega_0\in\Omega_\kappa$, otherwise the NLD wave with  
$\omega_0$ is unstable, \ie $\omega_0\in(0,1)\setminus \Omega_\kappa$.
For the sake of confidence in our results, $t_{fin}$ should be long enough,
and we choose $t_{fin}=40000$ in this work.

\begin{figure}
\centering
\includegraphics[width=8cm]{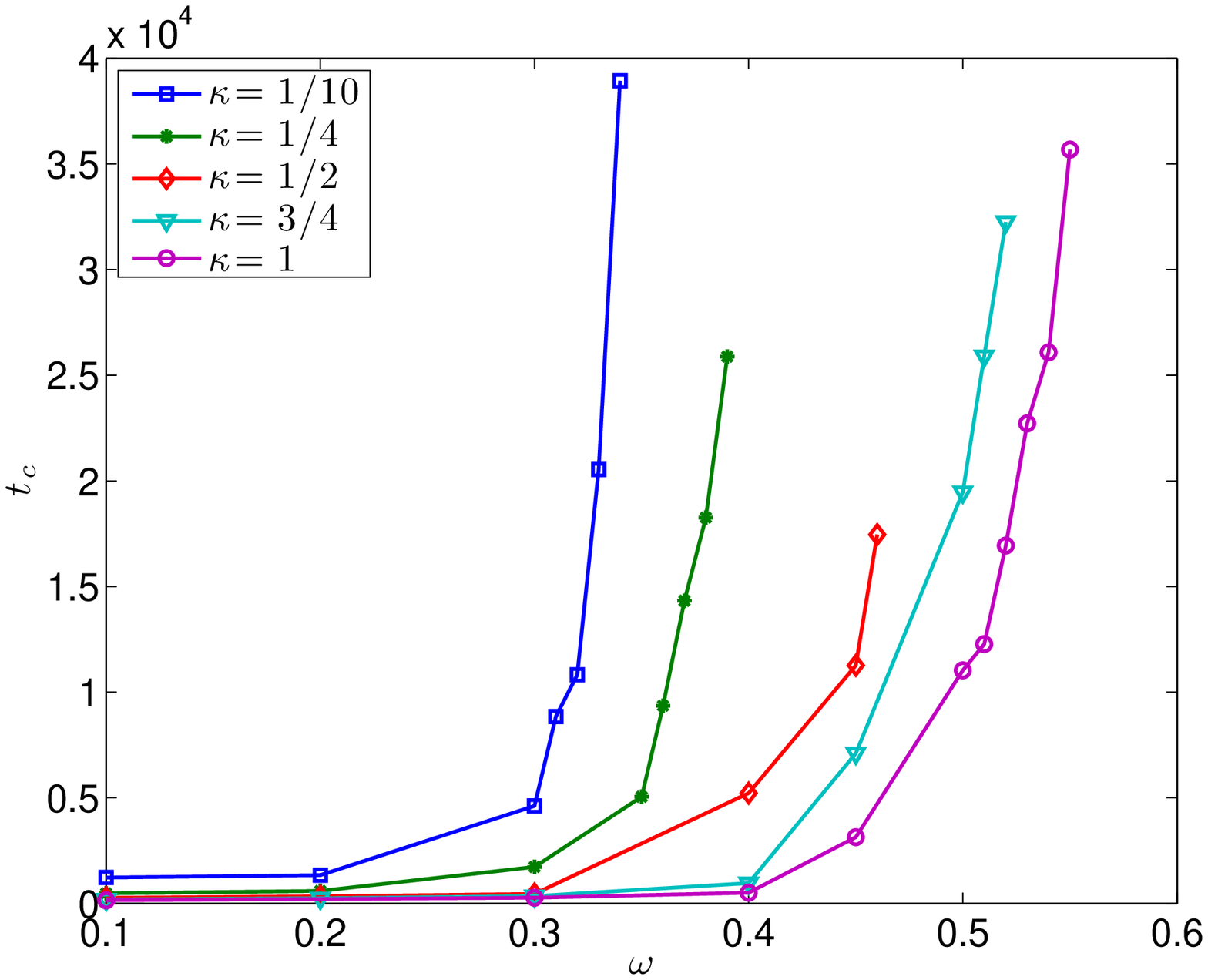}
\caption{\small Plots of the instant $t_c$ against the frequency $\omega$ for $0<\kappa\leq 1$.}
\label{fig:k1_omega}
\end{figure}

Our numerical simulations reveal that $\Omega_1 = [0.56,1)$. 
When the frequency approaches $0.56$ (the lower end of $\Omega_1$),
the instant of instability $t_c$ increases exponentially,
see Fig.~\ref{fig:k1_omega}. 

\begin{table}
\centering
\caption{\small Instants of time at which the monitored quantities ($q$, $\text{err}_{\infty}$, 
$\text{err}_{2}$, 
$\mathcal{V}_{P}$,       
$\mathcal{V}_{E}$,  
$\mathcal{V}_{Q}$) 
become larger than a given tolerance $\epsilon$ ($=1.00$E-$03$ here). The computational domain is $[-L,L]$ and five different lengths are tested.
Here $\tau=0.025$.}
\begin{tabular}{|c|c|c|c|c|c|c|}
\hline
$L$   & $q$ & $\text{err}_{\infty}$   &   $\text{err}_{2}$  & $\mathcal{V}_{P}$ &       $\mathcal{V}_{E}$ &  $\mathcal{V}_{Q}$\\
\hline
& \multicolumn{6}{|l|}{\textit{Two-hump wave with $\kappa=1$ and $\omega=0.1$}}\\
\hline
50    & 147 & 121 & 121 & 135 & 131 &  \\
\hline
75    & 146 & 122 & 122 & 135 & 132 &  \\
\hline
100   & 146 & 122 & 122 & 134 & 132 &  \\
\hline
125   & 146 & 122 & 120 & 135 & 139 &  \\
\hline
150   & 145 & 122 & 122 & 133 & 132 &  \\
\hline
& \multicolumn{6}{|l|}{\textit{One-hump wave with $\kappa=1$ and $\omega=0.5$}}\\
\hline
50    & 7373  & 6585 & 6614 & 6580 & 6601 & 6921 \\
\hline
75    & 9552  & 8728 & 8724 & 8720 & 8876 & 9177 \\
\hline
100   & 11036 & 9935 & 9937 & 9930 & 9930 & 10412 \\
\hline
125   & 12905 & 11673 & 11670 & 11672 & 11670 & 12183 \\
\hline
150   & 14641 & 13561 & 13560 & 13560 & 13560 & 14104 \\
\hline
\end{tabular}
\label{tab:Ldependent}
\end{table}

\subsection{$0<\kappa<1$}

\begin{figure}
\centering
\includegraphics[width=8cm]{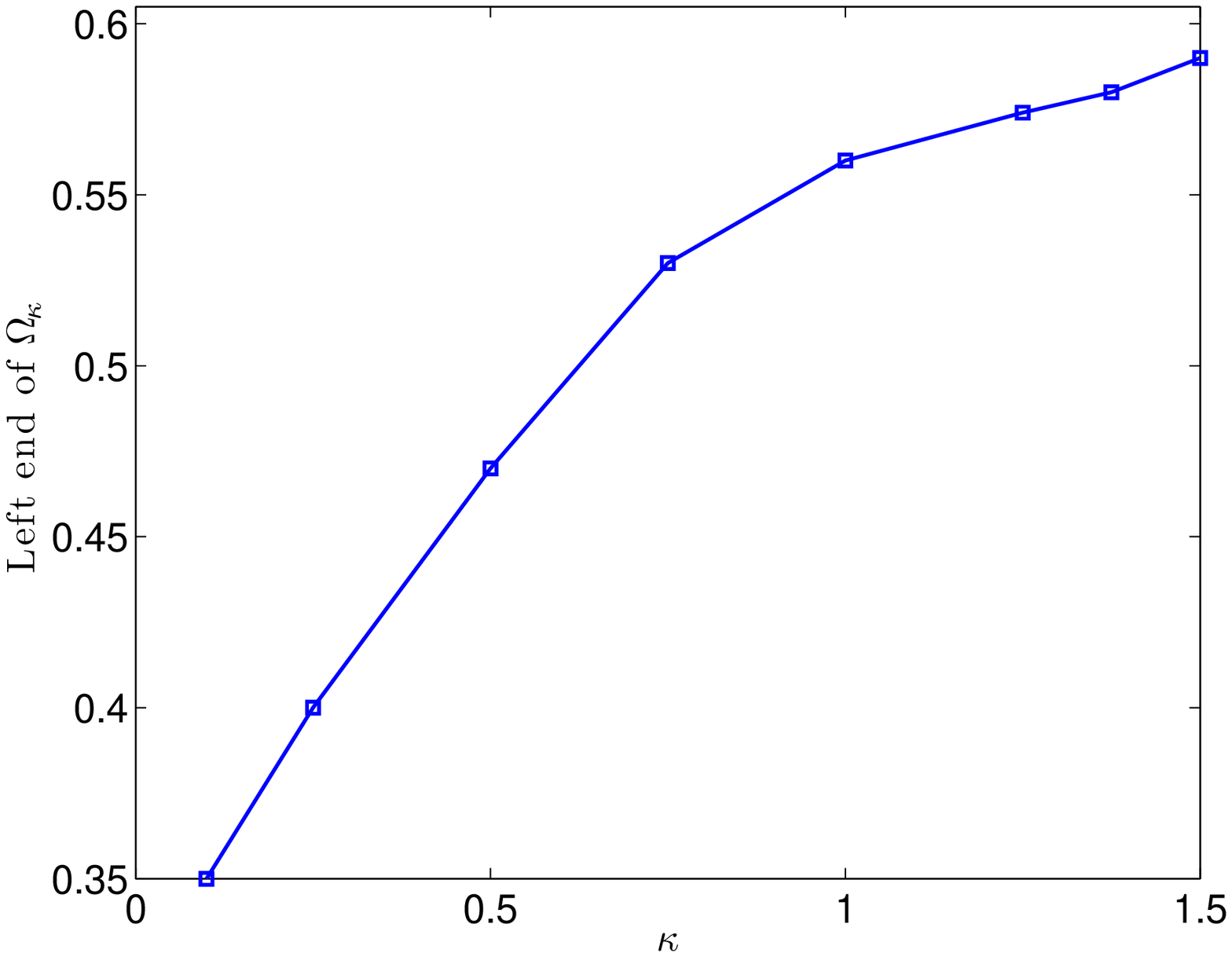}
\caption{\small Plot of the left end of the stable region $\Omega_\kappa$ for $0<\kappa\leq 3/2$.}
\label{fig:omega_c_kappa}
\end{figure}

For $\kappa\in(0,1)$, we find the stable region $\Omega_\kappa$ for $\omega$ as
follows:  $\Omega_{{1}/{10}} = [0.35,1)$, $\Omega_{{1}/{4}} = [0.40,1)$,
$\Omega_{{1}/{2}} = [0.47,1)$ and $\Omega_{{3}/{4}} = [0.53,1)$,
all of which are left-closed and right-open intervals 
with the same right end of $1$. Moreover, it is observed that 
the left end of $\Omega_\kappa$ increases monotonically as $\kappa$ increases from $0$ to $1$ and the limit is about $0.60$ for larger values of $\kappa$,  see Fig.~\ref{fig:omega_c_kappa}.

\subsection{$1<\kappa<2$}

\begin{figure}
\centering
\includegraphics[width=8cm]{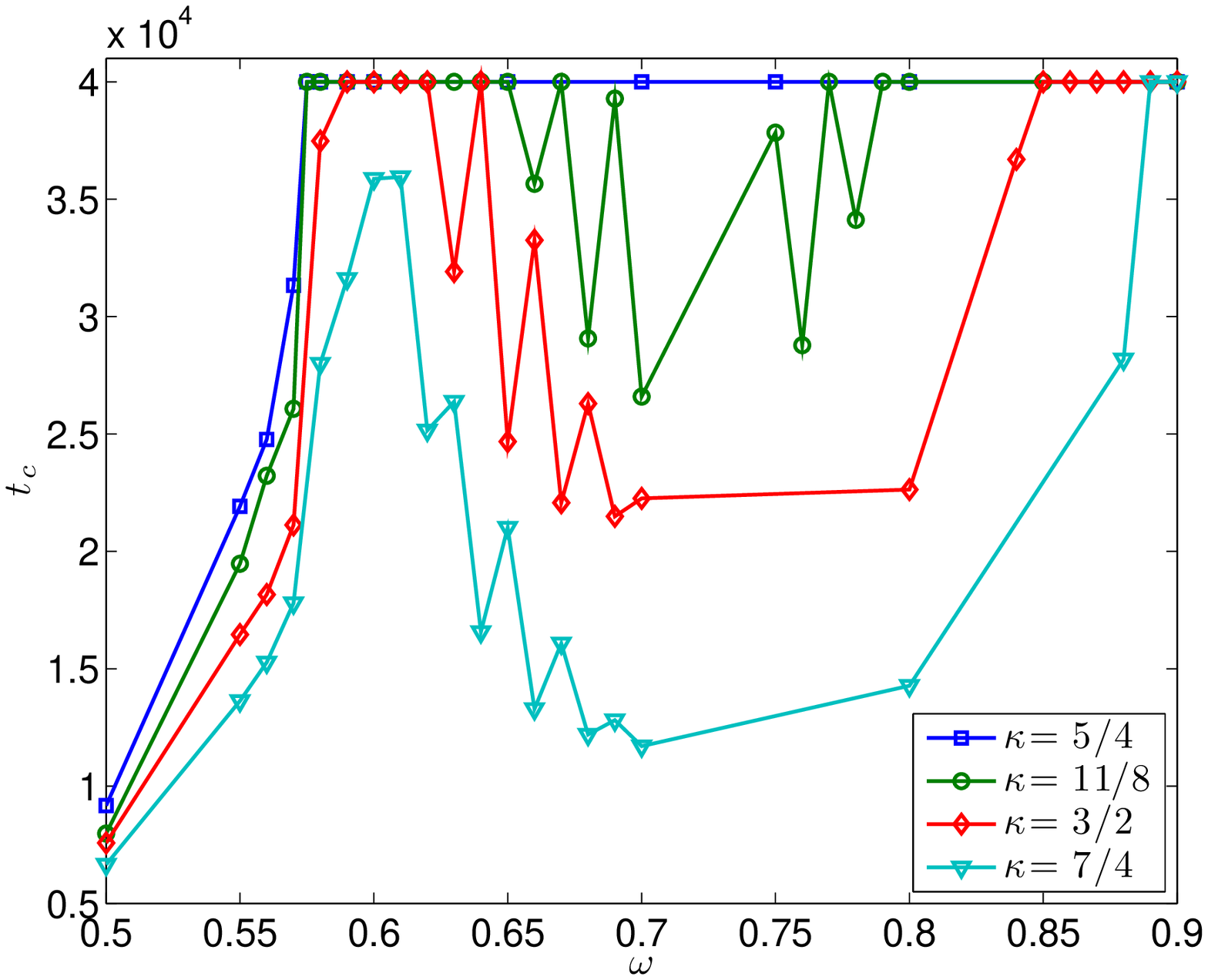}
\caption{\small Plot of $t_c$ vs. $\omega$ for $1<\kappa<2$. For the NLD waves in the stable region, $t_c$ is not available and we use $t_{fin}=40000$ instead. That is, the flat parts of the curve with a value of $40000$  correspond  
to the waves in the stable region.}
\label{fig:omega_c_kl2}
\end{figure}

For $\kappa\in(1,2)$, we find two types of stable region $\Omega_\kappa$:
the first type is a left-closed and right-open interval with 
the left end around $0.60$ and the right end at $1$,
\eg $\Omega_{{5}/{4}} = [0.58,1)$ and $\Omega_{{7}/{4}} = [0.89,1)$;
the second type consists of two disjoint intervals, \eg 
$\Omega_{{11}/{8}} = [0.58,0.67]\cup[0.77,1)$ 
and $\Omega_{{3}/{2}} = [0.59,0.64]\cup[0.85,1)$.
In Fig.~\ref{fig:omega_c_kl2},
we plot $t_c$ against $\omega$ for $1<\kappa<2$,
where $t_c$ is not available for the stable NLD waves and we use $t_{fin}=40000$ instead. That is, the flat part of the curve with a value of $40000$  corresponds to the waves in the stable region. 
It can be easily observed there that:
When the exponent power $\kappa$ is slightly larger than $1$, 
we have a large stable region of the first type;
when we keep increasing $\kappa$, 
this big stable region is divided into two small intervals located around the left end and the right end, respectively,  
which form together the stable region of the second type,
one closed interval with the left end around $0.60$
and the other left-closed and right-open interval with the right end of $1$;
when $\kappa$ approaches $2$,
the small interval around $0.60$ disappears
and then we have again the stable region of the first type 
but of much shorter length. 
As the frequency approaches the left end of $\Omega_\kappa$,
the instant of the instability $t_c$ increases exponentially.
In the case of stable region of the second type,
$t_c$ for the unstable NLD waves with 
the frequency $\omega$ between the two disjoint intervals 
oscillates in $\omega$ and decreases monotonically in $\kappa$
for a given frequency.

\subsection{$\kappa\geq 2$}

\begin{figure}
\centering
\includegraphics[width=8cm]{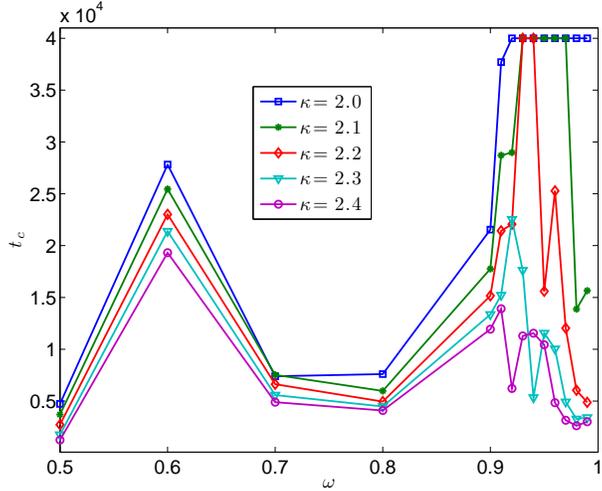}
\caption{\small Plot of $t_c$ vs. $\omega$ for $\kappa\geq 2$. For the NLD waves in the stable region, $t_c$ is not available and we use $t_{fin}=40000$ instead. That is, the flat parts of the curves with a value of $40000$  correspond to the waves in the stable region.}
\label{fig:omega_c_kb2}
\end{figure}

For $\kappa\geq 2$, 
the stable region exists only for $\kappa$ slightly larger than as well as equal to $2$, \eg $\Omega_{2} = [0.92,1)$, 
$\Omega_{2.1} = [0.93,0.97]$
and $\Omega_{2.2} = [0.93,0.94]$. For larger $\kappa$, 
the NLD waves are unstable for all $\omega\in(0,1)$,
\eg $\Omega_{2.3}=\Omega_{2.4}=\varnothing$.
In Fig.~\ref{fig:omega_c_kb2},
we plot $t_c$ against $\kappa$
and see that 
the instant of instability $t_c$ increases exponentially
as $\omega$ approaches the left end of $\Omega_\kappa$,
and decreases monotonically in $\kappa$ 
for a given frequency.

\subsection{Discussion}

According to the discovered stable region $\Omega_\kappa$ for $\kappa>0$
and Fig.~\ref{fig:profile}, 
we can conclude that all stable NLD waves are of one-hump profile,
which gives a positive answer to the conjecture raised in  
\cite{NLDE,XuShaoTangWei2013},
\ie the NLD waves of two-hump structure are unstable.
This is also in accordance with numerical observations in 
\cite{ShaoTang2005} which imply that 
the two-hump NLD solitary waves may collapse during scattering (\ie after collision they stop being solitary waves), whereas the collapse phenomena cannot be generally observed in collisions of the one-hump NLD solitary waves.

\begin{figure}
\centering
\includegraphics[width=8cm]{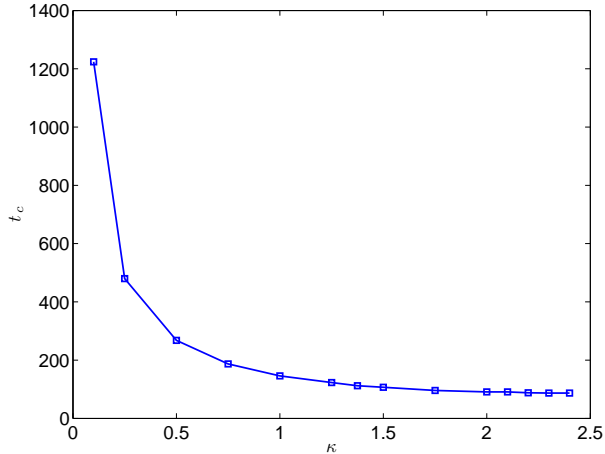}
\caption{\small Plot of $t_c$ against $\kappa$ for unstable NLD waves 
with $\omega=0.1$.}
\label{fig:t_c_kappa}
\end{figure}

When the exponent power $\kappa$ (denoting the strength of nonlinearity) 
increases, the stable region $\Omega_\kappa$ narrows.
For a given $\omega$ in the unstable region, 
the moment of instability $t_c$ decreases 
monotonically with increasing $\kappa$, see \eg Figs.~\ref{fig:omega_c_kl2} and \ref{fig:omega_c_kb2}. Particularly, for $\omega=0.1$,
we find that $t_c$ is inversely proportional to $\kappa$,
see Fig.~\ref{fig:t_c_kappa}.

\section{Summary} \label{sec6}

In this paper we reviewed various variational methods that had been put forward to determine possible criteria for the exact solitary wave solutions to the NLD equation to be unstable. We showed that these methods yield inconsistent results (in contrast to the NLS equation for which the results of all these methods agree):
The arguments of Bogolubsky suggested that for $\omega$ less than a critical value $\omega_B \approx 0.7$, which is practically independent of $\kappa$, the solitary waves should be unstable to slight changes in $\omega$ for fixed charge Q. An argument based on scale transformations suggested that the solitary wave solutions are unstable for all $\kappa > 1$. The Vakhitov-Kolokolov criterion suggested that for $\kappa < 2$ all solitary waves are stable and for $\kappa > 2$ there is a region of $\omega$ below a curve $\omega(\kappa)$    where the solitary waves are suggested to be stable. As the above suggestions yielded inconsistent results, we performed extensive numerical simulations in order to determine the stability regions $\Omega_{\kappa}$  for $\omega$. For $0 < \kappa < 1$ the stability regions are left-closed and right-open intervals with the same right end of 1, while the left end increases with $\kappa$.  For $\kappa = 1$ the stability interval is $[0.56,1)$. For $1 < \kappa < 2$ we find two types of $\Omega_{\kappa}$: The first one is a left-closed and right-open interval with the left end around $0.60$ and the right end at $1$. The second type consists of two disjoint intervals. For $\kappa =2$ there is a stable region just below $1$. For $\kappa > 2$ a very narrow stable region exists only for $\kappa$ slightly larger than $2$.  
For $0 < \kappa < 1$ the time $t_c$ when an instability sets in, increases exponentially with $\omega$ while the stable region is approached.  For 
$1 < \kappa < 2$, $t_c$ is a very complicated function of $\omega$ in the instability regions and $t_c$ decreases monotonically with increasing $\kappa$.
The stability of the solitary waves depends on their profile, i.e. on the shape of the charge density as a function of $x$. All stable waves have a one-hump profile, but not all one-hump waves are stable. All waves with two humps are unstable. 
An open issue is the study of collisions of NLD solitary waves with different 
$\kappa$ values.

\begin{acknowledgments}
This work was performed in part under the auspices of the United States Department of Energy. The authors would like to thank the Santa Fe Institute for its hospitality during the completion of this work. We also thank Prof. Comech for his useful comments on a draft of this paper. 
S.S. acknowledges financial support from the National Natural Science
Foundation of China (No.~11101011) and the Specialized
Research Fund for the Doctoral Program of Higher Education
(No.~20110001120112).
N.R.Q. 
acknowledges financial support from the Humboldt Foundation through Research Fellowship for Experienced Researchers SPA 1146358 STP and by the MICINN through 
FIS2011-24540, and by Junta de Andalucia under Projects No. FQM207, No. 
P06-FQM-01735, and No. P09-FQM-4643.  F.G.M. acknowledges the 
hospitality of the Mathematical Institute of the University of Seville (IMUS) and of the Theoretical 
Division and Center for Nonlinear Studies at Los Alamos National Laboratory, financial 
support by the Plan Propio of the University of Seville, and by the MICINN through FIS2011-24540. A.K. acknowledges financial support from Department of Atomic Energy,
Government of India through a Raja Ramanna Fellowship. 

\end{acknowledgments}

\end{document}